\documentclass[aip,jcp,preprint,superscriptaddress,floatfix]{revtex4-1}

\usepackage[utf8]{inputenc}
\usepackage{color}
\usepackage{graphicx}
\usepackage{mathtools}					

\newcommand{\ket}[1]{\left|#1\right>}
\newcommand{\bra}[1]{\left<#1\right|}
\newcommand{\kPsi}{\ket{\Psi}}						
\newcommand{\kPhi}{\ket{\Phi}}						
\newcommand{\allStates}{k_1,\ldots,k_L}	
\newcommand{\allStatesopp}{k_1^{\prime},\ldots,k_L^{\prime}} 
\newcommand{\alldim}{a_1,\ldots,a_{L-1}}                     
\newcommand{\alldimp}{a_1^{\prime},\ldots,a_{L-1}^{\prime}}  
\newcommand{\alldimop}{b_1,\ldots,b_{L-1}}                     
\newcommand{\ONvec}{\ket{k_1 \ldots k_L}}	
\newcommand{\ONvecb}{\bra{k_1^{\prime} \ldots k_L^{\prime}}}    
\newcommand{\ONstring}{k_1 \ldots k_L}		
\newcommand{\qcm}[1]{\textsc{QCMaquis}}

\begin{document}

\title{A nonorthogonal state-interaction approach for matrix product state wave functions}

\author{Stefan Knecht}
\email{stefan.knecht@phys.chem.ethz.ch}
\affiliation{ETH Z\"urich, 
Laboratorium f{\"u}r Physikalische Chemie, 
Vladimir-Prelog-Weg 2, 8093 Z\"urich, Switzerland}

\author{Sebastian Keller}
\affiliation{ETH Z\"urich, 
Laboratorium f{\"u}r Physikalische Chemie, 
Vladimir-Prelog-Weg 2, 8093 Z\"urich, Switzerland}

\author{Jochen Autschbach}
\affiliation{University at Buffalo, State University of New York, 
Department of Chemistry, 
New York 14260-3000, United States of America}

\author{Markus Reiher}
\affiliation{ETH Z\"urich, 
Laboratorium f{\"u}r Physikalische Chemie, 
Vladimir-Prelog-Weg 2, 8093 Z\"urich, Switzerland}

\begin{abstract}

  We present a state-interaction approach for matrix product state
  (MPS) wave functions in a nonorthogonal molecular orbital basis.
  Our approach allows us to calculate for example transition and
  spin-orbit coupling matrix elements between arbitrary electronic
  states provided that they share the same one-electron basis
  functions and active orbital space, respectively.  The key element
  is the transformation of the MPS wave functions of different states
  from a nonorthogonal to a biorthonormal molecular orbital basis
  representation exploiting a sequence of non-unitary transformations
  following a proposal by Malmqvist (\textit{Int. J. Quantum Chem.}
  \textbf{30}, 479 (1986)).  This is well-known for traditional
  wave-function parametrizations but has not yet been exploited for
  MPS wave functions.

\end{abstract}

\maketitle

\section{Introduction}\label{Introduction}

The calculation of electronic and vibronic transition matrix elements
between electronic states of the same or different spin and/or spatial
symmetry is a ubiquitous task in the modeling of photochemical and
photophysical processes.  Prime examples include the modeling of
non-adiabatic dynamics processes \cite{barb11,mai15} as well as
light-induced excited spin-state trapping phenomena
\cite{suau09,chan10}.  The theoretical description of these processes
builds upon the calculation of intersystem crossing rates
\cite{mari12} which, besides electronic and vibronic coupling
elements, requires the evaluation of spin-orbit (SO) coupling (SOC)
matrix elements.  Similarly, calculating magnetic properties
\cite{kaup04} such as molecular g-factors and electron-nucleus
hyperfine coupling, which are central parameters in electron
paramagnetic resonance (EPR) spectroscopy, requires spin-orbit coupled
wave functions \cite{gerl75,Autschbach:2015a}.  To this end,
correlated two- and four-component \textit{ab initio} wave function
\cite{bolv06,gany12,vad13,Autschbach:2014o}, and density functional
theory approaches \cite{vanl97,repi10,verm13}. In the
present study we focus on EPR g-tensors for testing purposes, but it
should be noted that the underlying novel method of gaining access to
wave functions that include the effects from SOC has a vast range of
applications.

While it is possible to treat SOC variationally, a considerable number
of two-step correlated wave function approaches for the calculation of
molecular g-factors were developed over the past decades (see, for
example,
Refs.~\citenum{lush96,brow03,nees05,bolv06,vanc07,nees07a,tatc09,roem15,sayf16}\
and literature citations in these works and in
Refs.~\citenum{Autschbach:2015a,bolv06,gany12,vad13,vanl97,repi10,verm13}). In
these schemes, the calculation of a number of non- or
scalar-relativistic many-particle spin-free states that are
eigenfunctions of the spin-squared operator $S^2$, is decoupled from a
subsequent perturbative or variational mixing of the latter through
the SO coupling operator to obtain SO coupled many-electron wave
functions (e.g. by diagonalization ``state-interaction'').  It is
straightforward to calculate properties such as g-factors in the basis
of the eigenstates of the SO operator subsequently.  Appealing
features of the two-step approaches are that valuable insight into
contributions of each (ground or excited) spin-state to the g-tensor
is gained, and that the underlying wave function basis is spin-adapted.
The price to pay is the need to calculate a sufficient number of
spin-free states to interact, which can amount up to several hundred
states to achieve convergence for (heavy-element containing) molecules
where electron correlation and spin-orbit coupling contributions can
be of similar order of magnitude.

Open shell electronic structures are often governed by strong electron
correlation effects.  In this context, multiconfigurational methods
are the preferred methods of choice \cite{szal12,roca12}\ which
typically split electron correlation into a static and a dynamic
contribution.  However, such a separation requires careful attention
\cite{stei16b}. A well-established approach to handle static
correlation is the complete active space self-consistent field
(CASSCF) ansatz \cite{olse11} which requires to select a tailored
number of (partially occupied) active orbitals. The selection of
active orbitals is a tedious procedure but can be automatized
\cite{stei16a,stei16b}.  Since the computational cost of traditional
CASSCF scales exponentially with the number of active orbitals and
electrons, tractable active orbital spaces are presently limited to
about 18 electrons in 18 orbitals \cite{aqui15}.  These limitations can
be overcome by resorting to the density matrix renormalization group
(DMRG) approach \cite{whit92,whit93,scho05,scho11} in quantum
chemistry
\cite{lege08,chan08,mart10,mart11,chan11a,wout14,kura14a,yana15,szal15,knec16,chan16}
which, in combination with a self-consistent-field orbital
optimization ansatz (DMRG-SCF) \cite{zgid08,ghos08}, is capable of
approximating CASSCF wave functions to chemical accuracy with merely a
polynomial scaling.  DMRG-SCF therefore allows to handle much larger
active orbital spaces that can boldly surpass the CASSCF limit.  To
account in addition for spin-orbit coupling in a DMRG framework,
variational SO approaches \cite{knec14}\ as well as two-step approaches
\cite{roem15,sayf16}\ based on spin-free DMRG wave functions have been
reported recently. 

In this work, we present a generalized state-interaction approach for
\textit{nonorthogonal} spin-free matrix product state (MPS) wave
functions which enables the evaluation of arbitrary one- and
two-particle transition matrix elements as well as SO coupling matrix
elements.  Diagonalization of the SO Hamiltonian matrix, for instance,
yields spin-orbit coupled wave functions as linear combinations of the
uncoupled, spin-pure MPS states. The latter can (but do not have to)
be obtained as results from one or several DMRG-SCF orbital
optimization calculations. This allows for utmost flexibility in the
individual DMRG-SCF steps as each state-specific or
  spin-specific state-averaged orbital optimization is given the
possibility to reflect potential relative differences in open-shell
occupancies. For example, transition metal as well as lanthanide and
actinide complexes often exhibit different $s$-\ and $d$-occupations
(transition metals) and $s$-, $d$-\ and $f$-occupations
(lanthanides/actinides) in ground- and electronically excited
electronic states of various spin symmetries.

However, a set of wave functions that were optimized individually
generally implies mutual nonorthogonality of the respective MO bases. Moreover, the MO bases may neither be orthogonal
to each other nor non-interacting which can, e.g., strongly affect the
calculation of transition moments between such electronic states
\cite{malm89}.  As first shown in a landmark paper by Malmqvist
\cite{malm86}, an elegant approach for the calculation of matrix
elements and transition density matrices is the transformation to a
biorthonormal basis for the bra and ket orbital basis of the
respective wave functions.  A change of the MO
basis to a biorthonormal orbital basis necessitates, however, not only
to transform all one- and two-electron integrals but to also
``counter-rotate'' the configuration basis of the wave function. For
configuration-interaction (CI)-type expansions, the rotations and
counter-rotations can be achieved by a sequence of single-orbital
transformations \cite{malm86,olse95}\ that require only one-electron
operations.  Subsequently, standard second-quantization algebra can be
exploited for the evaluation of overlap matrix elements as well as
arbitray one- and two-particle matrix elements between the states in
the biorthonormal basis. In a recent work, Olsen \cite{olse15}\ 
exploited the potential of the biorthonormal approach further to devise an efficient algorithm for CI and orbital optimization schemes 
based on nonorthogonal orbitals.  In contrast to previous nonorthogonal CI approaches 
(cf.~Ref.~\citenum{chen13b}), this newly proposed algorithm \cite{olse15}\ requires 
only the calculation of one- and two-particle reduced density matrices. 

In this paper, we derive the working equations to calculate one- and two-particle matrix 
elements between MPS wave functions that may be originally expressed in different, mutually 
nonorthogonal molecular orbital bases. Following the work of Malmqvist \cite{malm86}, 
the central element of our algorithm is the transformation of the bra and ket MPS wave functions 
to a biorthonormal basis representation. 
It is important to stress that the latter transformation is not needed 
if the MPS wave functions that are considered for state interaction 
share a common MO basis (cf.~Refs.~\citenum{roem15} and \citenum{sayf16}). 
After solving a generalized eigenvalue
equation of the form
\begin{equation}
{\rm \textbf{Hc}} = E {\rm \textbf{Sc}}\ ,
\end{equation}
with the Hamiltonian matrix \textbf{H} expressed in the basis of the DMRG-SCF 
wave functions and the overlap matrix \textbf{S}\ we obtain a set of fully orthogonal 
and non-interacting states as linear combinations of the DMRG-SCF 
wave functions with the expansion coefficients given by \textbf{c}.  

Since Malmqvist's approach \cite{malm86} assumes either a full CI expansion or, 
in general terms, a wave function expansion that is closed under de-excitation \cite{olse88,roos92,roos08a} 
we probe the closedness of our MPS wave function transformation by systematically increasing 
its numerical accuracy for a given active orbital space.

In Section \ref{sec:secqNOO}\ we briefly discuss the theoretical framework for a second quantization formalism 
based on nonorthogonal orbitals. In Section \ref{sec:mps-mpo}\ 
we introduce an algorithm to calculate expectation values for nonorthogonal wave functions 
in an MPS and matrix-product operator (MPO) representation of the wave function and 
operators \cite{kell15,kell16}, respectively, based on a nonunitary orbital transformation and demonstrate in Section \ref{sec:si-mps}\ 
how the latter can be exploited for a nonorthogonal MPS state-interaction (MPS-SI) \textit{ansatz}. 
Numerical examples for the calculation of g-factors for $f^1$-\ and $f^2$-type actinide 
complexes are presented in Section \ref{sec:secNUMEX}.  

\section{Second Quantization for Nonorthogonal Orbitals}\label{sec:secqNOO}

In this section, we briefly summarize the second quantization formalism for nonorthogonal orbitals \cite{mosh71,olse95,helg00,olse15}. 
In what follows, all quantities expressed in an orthonormal orbital basis are denoted by a tilde, 
those in the biorthonormal basis will be labeled by a bar 
whereas quantities with no extra label refer to a nonorthogonal orbital basis. 

Assuming an arbitrary set of $N$\ (linearly independent) MOs $\mathbf{\boldsymbol{\varphi}} = \{\varphi_p\}$ with the general 
overlap matrix ${\rm \bf S} = \left\{S_{pq}\right\}$\
\begin{equation}\label{eq:metric_noo}
 {S}_{pq}\ = \left<\varphi_p\left.\right|\varphi_q\right> = \int_{}^{}\ {\rm d\textbf{r}}\ \varphi_p^{\ast}({\rm \textbf{r}}) \varphi_q({\rm \textbf{r}})\ ,
\end{equation} 
we can define a new set of orbitals $\mathbf{\tilde{\boldsymbol{\varphi}}} = \{\tilde{\varphi_p}\}$\ that form an orthonormal 
basis by applying a L\"owdin symmetric orthogonalization \cite{lowd50} 
\begin{equation}\label{eq:trapreORB}
\mathbf{\tilde{\boldsymbol{\varphi}}} = \mathbf{\boldsymbol{\varphi}}\ {\rm \bf S}^{-1/2}\ .
\end{equation}
For an element of the overlap matrix in the new basis then holds
\begin{equation}
\tilde{S}_{pq} = \left({\rm \bf S}^{-1/2 \dagger}{\rm \bf S}{\rm \bf S}^{-1/2}\right)_{pq} = \delta_{pq}\ ,
\end{equation}
and consequently the new orbital set $\{\tilde{\varphi}\}$ is orthonormal. The corresponding set of creation (annihilation) operators $\{{\tilde{a}^{\dagger}_{p\sigma}}\}$ ($\{{\tilde{a}^{}_{p\tau}}\}$) for spin orbitals, $\tilde{\varphi}_{p}(\textbf{r})\sigma(m_s)$, in the new basis is related to the set of creation (annihilation) operators $\{{{a}^{\dagger}_{p\sigma}}\}$ ($\{{{a}^{}_{p\tau}}\}$) for spin orbitals of the original (nonorthogonal) basis \cite{helg00},
\begin{eqnarray}\label{eq:trapreCRE}
{\tilde{a}^{\dagger}_{p\sigma}} & = & \sum_{q}^{} {{a}^{\dagger}_{q\sigma}}\ {S}^{-1/2}_{qp}\ ,  \\
{\tilde{a}^{}_{p\sigma}} & = & \sum_{q}^{} {{a}^{}_{q\sigma}}\ {S}^{-1/2}_{pq}\ .
\end{eqnarray} 
Note that 
the creation $\{{\tilde{a}^{\dagger}_{p\sigma}}\}$\ and annihilation $\{{\tilde{a}^{}_{p\sigma}}\}$\ operators satisfy the well-known anticommutation rules. 

Expressing the creation (annhihilation) operators of the nonorthogonal basis in terms of the operators defined in the 
orthonormal basis yields \cite{helg00}
\begin{eqnarray}\label{eq:nonortCRE}
{{a}^{\dagger}_{p\sigma}}  &= &\sum_{q}^{} {\tilde{a}^{\dagger}_{q\sigma}}\ {S}^{1/2}_{qp}\ , \\
\label{eq:nonortANN}
{{a}^{}_{p\sigma}}  &= & \sum_{q}^{} {\tilde{a}^{}_{q\sigma}}\ {S}^{1/2}_{pq}\ .
\end{eqnarray} 
Inserting Eq.~(\ref{eq:nonortCRE}) into the definition of the anticommutator, it is easy to verify that 
two creation operators in the nonorthogonal basis anticommute as it is the case in the orthonormal basis,
\begin{equation}\label{eq:anticomm_noo}
\{{a}^{\dagger}_{p\sigma}, {a}^{\dagger}_{q\tau}\} = \sum_{{\substack{rs\\ \sigma\tau}}}\ \{\tilde{a}^{\dagger}_{r\sigma}\ {S}^{1/2}_{rp}, \tilde{a}^{\dagger}_{s\tau}\ {S}^{1/2}_{sq}\}\ = \sum_{{\substack{rs\\ \sigma\tau}}}\ {S}^{1/2}_{rp} {S}^{1/2}_{sq} \underset{= 0}{\underbrace{\{\tilde{a}^{\dagger}_{r\sigma}, \tilde{a}^{\dagger}_{s\tau}\}}}\ = 0 \ , 
\end{equation}
where we exploited in the second step the anticommutation of two creation operators for orthonormal spin orbitals. 
The same result holds for the anticommutation relation of two annihilation operators which is shown by 
Hermitian conjugation of Eq.~(\ref{eq:anticomm_noo}). 
The anticommutator between a creation and annihilation operator reads
\begin{equation}\label{eq:anticomm_CREANNnoo}
\{{a}^{\dagger}_{p\sigma}, {a}^{}_{q\tau}\} = \sum_{{\substack{rs\\ \sigma\tau}}}\ \{\tilde{a}^{\dagger}_{r\sigma}\ {S}^{1/2}_{rp}, \tilde{a}^{}_{s\tau}\ {S}^{1/2}_{qs}\}\ = \sum_{{\substack{rs\\ \sigma\tau}}}\ {S}^{1/2}_{rp} {S}^{1/2}_{qs} \underset{= \delta_{rs}\delta_{\sigma\tau}}{\underbrace{\{\tilde{a}^{\dagger}_{r\sigma}, \tilde{a}^{\dagger}_{s\tau}\}}}\ = \sum_{{\substack{r\\ \sigma\tau}}} {S}^{1/2}_{qr} {S}^{1/2}_{rp} \delta_{\sigma\tau}\ = {S}_{qp}\ \delta_{\sigma\tau}\ .
\end{equation}
Hence, whereas the anticommutator for a general pair $\tilde{a}^{\dagger}_{p\sigma},\tilde{a}^{\dagger}_{q\tau}$\ of creation- and annihilation operators reduces in the orthonormal orbital basis to 
\begin{equation}\label{eq:anticomm_CREANNstandard}
\{\tilde{a}^{\dagger}_{p\sigma}, \tilde{a}^{}_{q\tau}\} = \delta_{pq}\ \delta_{\sigma\tau},
\end{equation}
it depends on the (in general non-vanishing) overlap matrix element ${S}_{qp}$\ in the nonorthogonal case. 

As a consequence, the action of an annihilator $a_{q\sigma}$\ on a given occupation number vector (ONV) $\ket{\boldsymbol{k}_{}}$\
\begin{equation}\label{eq:onvDEF}
\ket{\boldsymbol{k}_{}} = \prod_{p=1}^{L} \left(a^{\dagger}_{p\sigma} \right)^{k_{p\sigma}} \ket{\rm {vac}} = \ket{k_{1} k_{2} \ldots k_{L}}\ ,
\end{equation}
with 
\begin{equation}
k_{p\sigma} = 
\left\{ 
\begin{array}{l@{\hspace{8pt}}l}
1 & \mbox{if}\ \varphi_{p}({\rm \textbf{r}}){\sigma(m_s)}\ {\rm occupied}\\
0 & \mbox{if}\ \varphi_{p}({\rm \textbf{r}}){\sigma(m_s)}\ {\rm unoccupied}
\end{array}
\right.
\end{equation}
leads to a sum of ONVs \cite{helg00}
\begin{equation}\label{eq:annihNOObasisaction}
a_{q\sigma} \ket{\boldsymbol{k}_{}} = \sum\limits_p^{L} (-1)^{\left[\sum\limits_{j=1}^{p-1} k_{j\sigma}\right]} {S}_{qp} k_{p\sigma} \ket{k_{1} k_{2} \ldots 0_{p} \ldots k_{L}} \ ,
\end{equation}
rather than to a single ONV as it would be the case for an orthonormal orbital basis. 
For the sake of completeness, we provide the proof for Eq.~(\ref{eq:annihNOObasisaction}) in the Appendix 
which exploits the anticommutator relation in Eq.~(\ref{eq:anticomm_CREANNnoo}).

In the last decades, considerable efforts were made to make nonorthogonal approaches such 
as valence-bond theory \cite{heit27,lowd91,vb_review2011,vanl13,chen13a,chen13b}\ and
nonorthogonal CI \cite{thom09,sund14,olse15}\ efficient. The latter suffer from an increasing computational 
complexity compared to a standard orthonormal formalism \cite{lowd55,pros68,welt76,dalg83,verb91,vanl91,koch93,amov97,vanl98}\ 
because of the nonorthogonal molecular orbital (MO) basis. 
In the framework of second quantization, the additional complexity can be attributed to 
the non-vanishing anticommutator in Eq.~(\ref{eq:anticomm_CREANNnoo}). 
For example, evaluating efficiently a matrix element of the type $\left<\Psi^{\rm}\right|\mathcal{O}\left|\Phi^{\rm}\right>$\ in a nonorthogonal
approach, where $\mathcal{O}$\ could be an arbitrary one-electron (two-electron) operator and $\ket{\Psi^{\rm}}$\ and
$\ket{\Phi^{\rm }}$\ are many-particle wave functions optimized for two different sets of MOs, respectively, 
requires a generalization \cite{lowd55,pros68,verb91}\ of the Slater-Condon rules \cite{slat29,cond30}.

In order to arrive at a formalism for evaluating matrix elements that closely resembles an orthonormal approach, 
the anticommutator in Eq.~(\ref{eq:anticomm_CREANNstandard})\ must vanish. 
To this end, it is useful to define a new orbital basis $\mathbf{\bar{\boldsymbol{\varphi}}} = \{\bar{\varphi}_{p}\} $\ 
through the transformation \cite{mosh71,doua92,olse95},
\begin{equation}\label{eq:trabioORB}
\mathbf{\bar{\boldsymbol{\varphi}}} = \mathbf{\boldsymbol{{\varphi}}}\  {\rm \bf S}^{-1}\ ,
\end{equation}
where $\{\bar{\varphi}_{p}\} $\ is referred to as the \emph{dual} of the nonorthogonal basis $\{\varphi_{p}\}$\ \cite{mosh71,doua92}. 
Moreover, $\{\bar{\varphi}_{p}\}$ and $\{\varphi_{p}\}$ are said to form a \emph{biorthonormal} system \cite{olse95}\ since we have 
\begin{equation}
\left<\bar{\varphi}_p\left.\right|\varphi_q\right> = \delta_{pq}\ ,
\end{equation}
and further, if both bases span the same space, they are jointly called a biorthonormal orbital basis.  
The definition of the biorthonormal creation operators follows from Eq.~(\ref{eq:trabioORB}) as 
\begin{eqnarray}
\label{eq:def_CREbio}
 \bar{a}^{\dagger}_{p\sigma} &=& \sum\limits_{r}^{} a^{\dagger}_{r\sigma} {S}_{rp}^{-1} \ .
\end{eqnarray}
To illustrate the biorthonormality, we consider the anticommutator for a biorthonormal creation operator with an annihilation operator in the original basis, 
\begin{eqnarray}\label{eq:anticomm_CREANbio}
\{\bar{a}^{\dagger}_{p\sigma}, {a}^{}_{q\tau}\} & = & \sum_{r}\ \{{a}^{\dagger}_{r\sigma}\ {S}^{-1}_{rp}, {a}^{}_{q\tau} \} = \sum_{r}\ {S}^{-1}_{rp} \underset{= {S}_{qr}\delta_{\sigma\tau}}{\underbrace{\{{a}^{\dagger}_{r\sigma}, {a}^{}_{q\tau}\}}} = \delta_{pq}\ \delta_{\sigma\tau}\ ,
\end{eqnarray}
which yields the standard anticommutation relation (cf. Eq.~(\ref{eq:anticomm_CREANNstandard})). 

\section{Matrix Product States and Matrix Product Operators}\label{sec:mps-mpo}

\subsection{Concepts}\label{sec:mps-mpo-concepts}

We briefly introduce the concepts of expressing a quantum state as an MPS and 
a (Hermitian) operator as an MPO. Our notation follows the presentation of Ref.~\citenum{kell15}.  

Consider an arbitrary state $\kPsi$\ in a Hilbert space spanned by $L$\ spatial orbitals 
which we express as a linear superposition of ONVs $\ket{\boldsymbol{k}}$\ with the CI coefficients $c_{\ONstring}$\ as expansion coefficients 
\begin{equation}\label{eq:CI_wave_function}
\kPsi = \sum\limits_{{\boldsymbol{k}}} c_{\boldsymbol{k}} \ket{\boldsymbol{k}} = \sum\limits_{\allStates} c_{\ONstring} \ONvec\ ,
\end{equation}
where each local space is of dimension four corresponding to the 
basis states $k_l = \left|\uparrow\downarrow\right>, \left|\uparrow\right>, \left|\downarrow\right>, \left|0\right>$\ of a spatial orbital. 
In an MPS representation of $\kPsi$, we encode the CI 
coefficients $c_{\ONstring}$\ as a product of $m_{l-1}\times m_{l}$-dimensional matrices $M^{k_l} = \{M^{k_l}_{a_{l-1}a_l}\}$ 
\begin{align}
\kPsi &= \sum_{\allStates} \sum_{\alldim} M^{k_1}_{1 a_1} M^{k_2}_{a_1 a_2} \cdots M^{k_L}_{a_{L-1} 1} \ONvec = \sum_{\boldsymbol{k}} M^{k_1} M^{k_2} \cdots M^{k_L} \ket{\boldsymbol{k}}\ , \label{eq:MPS2}
\end{align}
where the last equality follows from collapsing the summation over 
the $a_l$\ indices (sometimes referred to as \emph{virtual indices} or \emph{bonds}) as matrix-matrix multiplications. 
Since the final contraction of the matrices $M^{k_l}$\ must yield the scalar coefficient $c_{\ONstring}$, 
the first and the last matrices are in practice 
$1\times m_1$-dimensional row and $m_{L-1}\times 1$-dimenisional column vectors, respectively. 
Allowing for the introduction of some maximum dimension $m$ for the matrices $M^{k_l}$, with $m$\ commonly 
referred to as \textit{number of renormalized block states} \cite{whit93}, is the central idea that facilitates a reduction of the exponentially 
scaling full CI \textit{ansatz} in Eq.~(\ref{eq:CI_wave_function}) to a polynomial-scaling MPS wave function \textit{ansatz}. 
For further details on the actual variational search algorithm for ground- and excited states in an MPS framework, 
we refer the reader to the review by Schollw\"ock \cite{scho11}\ and, for its formulation in a quantum chemical program 
package, for instance to our recent works \cite{kell15,kell16}. 

We may express an operator $\widehat{W}$\ in MPO form 
\begin{eqnarray}\label{eq:MPO}
	\widehat{\mathcal{W}} & = & \sum_{\allStates} \sum_{\allStatesopp} \sum_{\alldimop} W^{k_1 k_1^{\prime}}_{1 b_1} W^{k_2 k_2^{\prime}}_{b_1 b_2} \cdots W^{k_L k_L^{\prime}}_{b_{L-1} 1} 
	\ONvec \ONvecb\nonumber\\
	         & = & \sum_{\boldsymbol{k} \boldsymbol{k^{\prime}}} W^{k_1 k_1^{\prime}} W^{k_2 k_2^{\prime}} \cdots W^{k_L k_L^{\prime}}  \ket{\boldsymbol{k}}\bra{\boldsymbol{k^{\prime}}} \equiv \sum_{\boldsymbol{k} \boldsymbol{k^{\prime}}} w_{\boldsymbol{k} \boldsymbol{k^{\prime}}} \ket{\boldsymbol{k}}\bra{\boldsymbol{k^{\prime}}}\ ,
\end{eqnarray}
with the incoming and outgoing physical states $k_l$\ and $k_l^{\prime}$\ and the virtual indices $b_{l-1}$\ and $b_l$. 
In analogy to Eq.~(\ref{eq:MPS2}), we may recognize the summation over pairwise matching indices $b_l$\ as matrix-matrix 
multiplications which leads to the second line on the right-hand side of Eq.~(\ref{eq:MPO}). 
For practical applications, we rearrange the summations in Eq.~({\ref{eq:MPO}})\ and contract 
first over the local site indices $k_l k_l^{\prime}$\ 
\begin{equation}\label{eq:local_contract}
\widehat{W}_{b_{l-1} b_{l}} = \sum\limits_{k_l k_l^{\prime}} W^{k_l k_l^{\prime}}_{b_{l-1} b_l} \ket{k_l} \bra{k_l^{\prime}}\ ,
\end{equation}
which then yields 
\begin{equation}\label{eq:mpoMOD}
	\widehat{\mathcal{W}}  = \sum\limits_{\alldimop} W^{1}_{1 b_{1}} \cdots W^{l}_{b_{l-1} b_{l}} \cdots W^{L}_{b_{L-1} 1}\ .
\end{equation}
The local, operator-valued matrix representation introduced in Eq.~(\ref{eq:local_contract})\ 
is a central element for an efficient MPO-based implementation of the quantum-chemical DMRG approach \cite{kell15,kell16}\ 
that offers the same polynomial scaling as a ``traditional"\ (non-MPO) DMRG implementation.   

\subsection{MPO expectation values with nonorthogonal orbitals}\label{subsec:expMPSMPOnonoo}

The calculation of overlap matrix elements and expectation values for $N$-electron operators in an MPO framework based 
on an orthonormal basis 
common for the bra and ket states was outlined, for instance, in Refs. \citenum{scho11}\ and \citenum{kell15}, respectively. 
Here, we illustrate the complexity that results when the bra and ket states are expressed in 
two different MO bases which are in general not orthonormal.

Following the notation by Malmqvist and Roos \cite{malm89}, orbital basis superscripts ${\rm X}$\ and ${\rm Y}$\ denote 
the original (orthonormal) MO bases $\{\varphi^{\rm X}_{p}\}$\ and $\{\varphi^{\rm Y}_{p}\}$, respectively, 
whereas the superscripts ${\rm A}$\ and ${\rm B}$\ 
refer to different MO bases $\{\varphi^{\rm A}_{p}\}$\ and $\{\varphi^{\rm B}_{p}\}$, respectively, 
which have been manipulated in some way and are therefore in general not orthonormal.
We assume that the MPS wave functions are optimized with the same active orbital spaces and a common atomic orbital basis set. 
The latter restrictions can, however, be lifted \cite{olse95,malm02}.

\subsubsection{General considerations}\label{sec:general}

Let $\kPsi$\ and $\kPhi$\ denote two MPS wave functions based on the definition in Eq.~(\ref{eq:MPS2}),
\begin{eqnarray}
\kPsi &= & \sum_{\boldsymbol{k}^{\rm A}} \sum_{\alldim} M^{k_1^{\rm A}}_{1 a_1} M^{k_2^{\rm A}}_{a_1 a_2} \cdots M^{k_L^{\rm A}}_{a_{L-1} 1} \ket{\boldsymbol{k}^{\rm A}}\ ,\\
\kPhi &= & \sum_{\boldsymbol{k}^{\rm B}} \sum_{\alldimp} N^{k_1^{\rm B}}_{1 a_1^{\prime}} N^{k_2^{\rm B}}_{a_1^{\prime} a_2^{\prime}} \cdots N^{k_L^{\rm B}}_{a_{L-1}^{\prime} 1} \ket{\boldsymbol{k}^{\rm B}}\ ,
\end{eqnarray}
which are constructed from MO sets $\{\varphi^{\rm A}_{p}\}$\ and $\{\varphi^{\rm B}_{p}\}$, respectively. 

We start by considering the overlap $\left<\Phi\right|\Psi\left>\right.$ which can be written as \cite{helg00,scho11,kell15}
\begin{eqnarray}\label{eq:overlapMPS}
\left<\Phi\right|\Psi\left>\right. & = &\sum_{\boldsymbol{k}^{\rm A} \boldsymbol{k}^{\rm B}} \sum\limits_{\substack{\alldim \\ \alldimp}} \left(N^{k_L^{\rm B}\dagger}_{1 a_{L-1}^{\prime} } \cdots N^{k_2^{\rm B}\dagger}_{a_2^{\prime} a_1^{\prime} } N^{k_1^{\rm B}\dagger}_{a_1^{\prime} 1}\right) \left(M^{k_1^{\rm A}}_{1 a_1} M^{k_2^{\rm A}}_{a_1 a_2} \cdots M^{k_L^{\rm A}}_{a_{L-1} 1}\right) \left<\boldsymbol{k}^{\rm B}\right|\boldsymbol{k}^{\rm A}\left>\right. \nonumber\\
& = &  \sum\limits_{\boldsymbol{k}^{\rm A} \boldsymbol{k}^{\rm B}}\sum\limits_{\substack{\alldim \\ \alldimp}} \left(N^{k_L^{\rm B}\dagger}_{1 a_{L-1}^{\prime} } \cdots N^{k_2^{\rm B}\dagger}_{a_2^{\prime} a_1^{\prime} } N^{k_1^{\rm B}\dagger}_{a_1^{\prime} 1}\right)\left(M^{k_1^{\rm A}}_{1 a_1} M^{k_2^{\rm A}}_{a_1 a_2} \cdots M^{k_L^{\rm A}}_{a_{L-1} 1}\right) \nonumber \\
& & \qquad \qquad \qquad \times \delta_{N_{\boldsymbol{k}^{\rm B}} N_{\boldsymbol{k}^{\rm A}}}\det {\rm \textbf{S}}^{\boldsymbol{k}^{\rm B} \boldsymbol{k}^{\rm A}}\ ,
\end{eqnarray}
with $N_{\boldsymbol{k}^{\rm B}}$\ and $N_{\boldsymbol{k}^{\rm A}}$\ being the 
number of electrons comprised in the ONVs and $\det{\rm \textbf{S}}^{\boldsymbol{k}^{\rm B} \boldsymbol{k}^{\rm A}}$\ the 
determinant of the overlap matrix ${\rm \textbf{S}}^{\boldsymbol{k}^{\rm B} \boldsymbol{k}^{\rm A}}$\ of  
all overlap integrals between occupied orbitals of the bra and ket ONVs. 
Compared to the standard expression of the inner product of ONVs that are built 
from a common orthonormal basis $\{\tilde{\varphi}\}$, implying for example $\{\varphi^{\rm A}_{p}\} \equiv \{\varphi^{\rm B}_{p}\}$,\cite{helg00} 
\begin{equation}\label{eq:ONVoverlap}
\left<\boldsymbol{k}^{\rm B}\right|\boldsymbol{k}^{\rm A}\left>\right. = \prod_{p=1}^{L}\delta_{k_{p}^{\rm B}k_p^{\rm A}}\ ,
\end{equation}
the inner product in Eq.~(\ref{eq:overlapMPS})\ leads to a rather complex expression with a large 
number of non-zero terms. 
The reason for the latter is that in the general case of nonorthogonal orbitals the action of an annihilator 
on a given ONV, as illustrated in Eq.~(\ref{eq:annihNOObasisaction}), creates a sum of ONVs rather than a single ONV. 

Turning next to transition matrix elements $\left<{\Phi}\right|\mathcal{\hat{O}}\left|{\Psi}\right>$\ of an $N$-electron 
operator $\mathcal{\hat{O}}$\ given in MPO form (cf.\ Eq~(\ref{eq:MPO})), the expression in the case of nonorthogonal orbitals reads 
\begin{align}
\bra{\Phi}\mathcal{\hat{O}}\ket{\Psi} = & 
\sum\limits_{\boldsymbol{k}^{\rm A} \boldsymbol{k}^{\rm B}} \sum\limits_{\substack{\alldim \\ \alldimp}} \left(N^{k_L^{\rm B}\dagger}_{1 a_{L-1}^{\prime} } \cdots N^{k_2^{\rm B}\dagger}_{a_2^{\prime} a_1^{\prime} } N^{k_1^{\rm B}\dagger}_{a_1^{\prime} 1}\right) \nonumber \\ 
&   \times \left<\boldsymbol{k}^{\rm B}\right| \left[\  
\sum\limits_{\boldsymbol{k^{\rm B \prime}k^{\rm B \prime \prime}}}\ 
\sum\limits_{\boldsymbol{k^{\rm A \prime}k^{\rm A \prime \prime}}}\ 
\sum\limits_{b_1\ldots b_{L-1}}\  \left(O^{k_1^{\rm B \prime} k_1^{\rm A \prime}}_{1 b_1} O^{k_2^{\rm B \prime} k_2^{\rm A \prime}}_{b_1 b_2} \cdots O^{k_L^{\rm B \prime} k_L^{\rm A \prime}}_{b_{L-1} 1} \right) \right. \left.\left|{\boldsymbol{k}^{\rm B \prime}}\right>\left<{\boldsymbol{k}^{\rm B \prime \prime}}\right|\left.\boldsymbol{k}^{\rm A \prime \prime}\right>\left<{\boldsymbol{k}^{\rm A \prime}}\right| 
\vphantom{\sum\limits_{b_1\ldots b_{L-1}}} \right]  \nonumber \\
&  \times  \left(M^{k_1^{\rm A}}_{1 a_1} M^{k_2^{\rm A}}_{a_1 a_2} \cdots M^{k_L^{\rm A}}_{a_{L-1} 1}\right) \left|{\boldsymbol{k}^{\rm A}}\right>\ .
\label{eq:mps-epx-noo}
\end{align}
Note that, similarly to Eq.~(\ref{eq:overlapMPS}), the inner product $\left<{\boldsymbol{k}^{\rm B \prime \prime}}\right|\left.\boldsymbol{k}^{\rm A \prime \prime}\right>$\ 
of the ONVs\ will not simply reduce to a product of delta functions since the incoming and outgoing basis states 
of the operator are nonorthogonal. Moreover, the fermionic anticommutation which is encoded explicitly in the 
matrix representation of the creation and annihilation operators (see, in particular, Section IV in Ref. \citenum{kell15} for a detailed discussion) 
by using their Wigner-Jordan-transformed form \cite{jord28}\ will introduce additional overlap terms as becomes 
clear from the anticommutation rules in Eqs. (\ref{eq:anticomm_noo}) and (\ref{eq:anticomm_CREANNnoo}).

If, however, we assume that $\kPsi$\ and $\kPhi$\ are expressed in a common, orthonormal basis\ where Eq.~(\ref{eq:ONVoverlap})\ holds, 
Eq.~(\ref{eq:mps-epx-noo})\ will reduce to the well-known expression \cite{scho11,kell15},
\begin{eqnarray}\label{eq:expNORMAL}
\left<{\Phi}\right|\mathcal{\hat{O}}\left|{\Psi}\right> & = & 
\sum\limits_{\boldsymbol{k^{\rm A}} \boldsymbol{k^{\rm B}}} \sum\limits_{\substack{\alldim \\ \alldimp}} \left(N^{k_L^{\rm B}\dagger}_{1 a_{L-1}^{\prime} } \cdots N^{k_2^{\rm B}\dagger}_{a_2^{\prime} a_1^{\prime} } N^{k_1^{\rm B}\dagger}_{a_1^{\prime} 1}\right) \nonumber \\
& & \times \left<\boldsymbol{k}^{\rm B}\right| \left[\  
\sum\limits_{\boldsymbol{k^{\rm B \prime}k^{\rm B \prime \prime}}}\ 
\sum\limits_{\boldsymbol{k^{\rm A \prime}k^{\rm A \prime \prime}}}\ 
\sum\limits_{b_1\ldots b_{L-1}}\  \left(O^{k_1^{\rm B \prime} k_1^{\rm A \prime}}_{1 b_1} O^{k_2^{\rm B \prime} k_2^{\rm A \prime}}_{b_1 b_2} \cdots O^{k_L^{\rm B \prime} k_L^{\rm A \prime}}_{b_{L-1} 1} \right) \right.
\left.\left|{\boldsymbol{k}^{\rm B \prime}}\right>\underset{= \delta_{\boldsymbol{k}^{\rm B \prime \prime} \boldsymbol{k}^{\rm A \prime \prime}} = \mathbf{\rm I}}{\underbrace{\left<{\boldsymbol{k}^{\rm B \prime \prime}}\right|\left.\boldsymbol{k}^{\rm A \prime \prime}\right>}}\left<{\boldsymbol{k}^{\rm A \prime}}\right|
\vphantom{\sum\limits_{\boldsymbol{k^{\rm B \prime}k^{\rm B \prime \prime}}}\ 
	\sum\limits_{\boldsymbol{k^{\rm A \prime}k^{\rm A \prime \prime}}}\ 
	\sum\limits_{b_1\ldots b_{L-1}}} \right]  \nonumber \\
& & \times \left(M^{k_1^{\rm A}}_{1 a_1} M^{k_2^{\rm A}}_{a_1 a_2} \cdots M^{k_L^{\rm A}}_{a_{L-1} 1}\right) \ket{\boldsymbol{k}^{\rm A}}\ \nonumber\\
& = & 
\sum\limits_{\boldsymbol{k^{\rm A \prime}k^{\rm A}}}\ 
\sum\limits_{\boldsymbol{k^{\rm B \prime}k^{\rm B }}}\ 
\sum\limits_{b_1\ldots b_{L-1}}\
\sum\limits_{\substack{\alldim \\ \alldimp}}\ 
\left(N^{k_L^{\rm B}\dagger}_{1 a_{L-1}^{\prime} } \cdots N^{k_2^{\rm B}\dagger}_{a_2^{\prime} a_1^{\prime} } N^{k_1^{\rm B}\dagger}_{a_1^{\prime} 1}\right) 
\left(O^{k_1^{\rm B \prime} k_1^{\rm A \prime}}_{1 b_1} O^{k_2^{\rm B \prime} k_2^{\rm A \prime}}_{b_1 b_2} \cdots O^{k_L^{\rm B \prime} k_L^{\rm A \prime}}_{b_{L-1} 1} \right) \nonumber\\
& & \times \left(M^{k_1^{\rm A}}_{1 a_1} M^{k_2^{\rm A}}_{a_1 a_2} \cdots M^{k_L^{\rm A}}_{a_{L-1} 1}\right)
\underset{= \delta_{\boldsymbol{k^{\rm B}k^{\rm B \prime}}}}{\left<\right.\boldsymbol{k}^{\rm B}\left.\right|\boldsymbol{\rm k}^{\rm B\prime}\left.\right>} \underset{= \delta_{\boldsymbol{k^{\rm A\prime} k^{\rm A}}}}{\left<\right.\boldsymbol{k}^{\rm A\prime}\left.\right|\boldsymbol{k}^{\rm A}\left.\right>}\ \nonumber\\
& = & \sum\limits_{\substack{{k}_L^{\rm B} {k}_L^{\rm A}\\ a_{L-1}^{} a_{L-1}^{\prime} b_{L-1} }} N^{k_L^{\rm B} \dagger}_{1 a_{L-1}^{\prime}} O^{k_L^{\rm B} k_L^{\rm A}}_{b_{L-1} 1}\left(\cdots\sum\limits_{\substack{k_2^{\rm B} k_2^{\rm A}\\ a_1 a_1^{\prime} b_1}} N^{k_2^{\rm B}\dagger}_{a_2^{\prime} a_1^{\prime}} O^{k_2^{\rm B}k_2^{\rm A}}_{b_1 b_2}\left(\sum_{k_1^{\rm B} k_1^{\rm A}} N^{k_1^{\rm B} \dagger}_{a_1^{\prime} 1} O^{k_1^{\rm B}k_1^{\rm A}}_{1 b_{1}}M^{k_1^{\rm A}}_{1 a_1}\right) \right. \nonumber \\ 
 & & \times \left. M^{k_2^{\rm A}}_{a_1 a_2} \cdots\vphantom{\sum\limits_{\substack{{k}_L^{\rm B} {k}_L^{\rm A}\\ a_{L-1}^{} a_{L-1}^{\prime} b_{L-1} }} N^{k_L^{\rm B} \dagger}_{1 a_{L-1}^{\prime}} O^{k_L^{\rm B} k_L^{\rm A}}_{b_{L-1} 1}\left(\sum_{k_1^{\rm B} k_1^{\rm A}} N^{k_1^{\rm B} \dagger}_{a_1^{\prime} 1} O^{k_1^{\rm B}k_1^{\rm A}}_{1 b_{1}}M^{k_1^{\rm A}}_{1 a_1}\right)}\right)M^{k_L^{\rm A}}_{a_{L-1} 1}\ .
\end{eqnarray}
Note that the last equality follows from the original expression by regrouping the summations to minimize the operational cost \cite{kell15, scho11}. 

\subsubsection{Transformation of orbitals and MPS wave functions}\label{subsubsec:orb-mps-trafo}

To bring the expressions for the overlap and transition matrix elements,  
Eqs.~(\ref{eq:overlapMPS})\ and (\ref{eq:mps-epx-noo}), into a form that is comparable to the case for a 
common orthonormal basis requires the fulfillment of the biorthonormality condition \cite{mosh71,doua92} 
\begin{equation}\label{eq:biortho-cond}
\left<\bar{\varphi}^{\rm A}_p\right|\varphi^{\rm B}_q\left>\right. = \delta_{pq}\ .
\end{equation}
The price to pay is, however, that a transformation to a biorthonormal MO basis entails an additional transformation step 
of the wave function expansion parameters, in the present case the MPS tensors. 
This is most easily seen by inspection of Eq.~(\ref{eq:def_CREbio}) which defines a creation operator in the 
biorthonormal basis as a suitable linear combination of creation operators in the original basis. 
Because of the latter basis change of the creation (annihilation) operators the wave function expansion 
parameters need to be transformed as well. Moreover, as already pointed out by Malmqvist and Roos
\cite{malm89}, the objective is to find efficient wave function transformations from the original 
basis $\{\varphi^{\rm X}_{p}\}$ ($\{\varphi^{\rm Y}_{p}\}$) to the biorthogonal basis 
$\{\varphi^{\rm A}_{p}\}$ ($\{\varphi^{\rm B}_{p}\}$) \textit{while explicitly avoiding}\ any additional wave function optimization. 
Such an algorithm was proposed by Malmqvist for CI-type wave functions \cite{malm86}. 
Following up on Malmqvist's original idea, we outline in the following how such a transformation 
can be achieved for MPS-type wave function expansions by a sequence of single-orbital replacements. 
Similarly to CI-type expansions,
each step in the sequence is approximately twice as expensive to compute as the matrix-vector product
\begin{equation}\label{eq:htimesx}
y = {\widehat{\mathcal{T}}} x\ .
\end{equation}
for a one-electron-operator ${\widehat{\mathcal{T}}}$.

\paragraph{Orbital transformation}

Considering the biorthonormality condition (cf. Eq.~(\ref{eq:biortho-cond})), Malmqvist \cite{malm86}\ and 
Olsen \textit{et al.}\ \cite{olse95}\ showed that an LU-factorization of the inverse of the orbital overlap 
matrix $\left(\rm{\textbf{S}^{\rm XY}}\right)^{-1}$\ defined according to Eq.~(\ref{eq:metric_noo})\ 
with the bra (ket) orbitals in $\{\varphi^{\rm X}_{p}\}$ ($\{\varphi^{\rm Y}_{p}\}$) yields
\begin{equation}\label{eq:transmat}
\left(\rm{\textbf{S}^{\rm XY}}\right)^{-1} = {\rm \bf C}^{\rm YB} \left({\rm \bf C}^{\rm XA}\right)^{\dagger}\ ,
\end{equation}
where $\rm{\textbf{C}^{\rm XA}}$\ and $\rm{\textbf{C}^{\rm YB}}$\ are the desired orbital transformation matrices.  
They allow us to express the biorthonormal bases $\{{\varphi}^{\rm A}_{p}\}$\ and $\{{\varphi}^{\rm B}_{p}\}$\ in terms of 
the original orbital bases $\{{\varphi}^{X}_{p}\}$\ and $\{{\varphi}^{Y}_{p}\}$, respectively, \cite{malm86,olse95}
\begin{eqnarray}
{{\boldsymbol{\varphi}}^{\rm A}} & = & {{\boldsymbol{\varphi}}^{\rm X}}\ {\rm \bf C}^{\rm XA}\ , \label{eq:transbasis-A}\\
{{\boldsymbol{\varphi}}^{\rm B}} & = &{{\boldsymbol{\varphi}}^{\rm Y}}\ {\rm \bf C}^{\rm YB}\ \label{eq:transbasis-B}\ .
\end{eqnarray}
Expressing the orbital transformation in Eq.~(\ref{eq:transbasis-A})\ as a sequence of single orbital transformations 
the substitution of orbital $\varphi_1^{\rm X}$\ reads \cite{malm86,olse95} 
\begin{equation}
\varphi_1^{\rm X} = \varphi_1^{\rm A} t_{11} + \varphi_2^{\rm X} t_{21} + \varphi_3^{\rm X} t_{31} + \ldots\ ,
\end{equation}
and for orbital $\varphi_2^{\rm X}$\ 
\begin{equation}
\varphi_2^{\rm X} = \varphi_1^{\rm A} t_{12} + \varphi_2^{\rm A} t_{22} + \varphi_3^{\rm X} t_{32} + \ldots\ ,
\end{equation}
and so forth. 
Collecting the unknown parameters $t_{mn}$\ in a matrix $\rm{\textbf{t}}$\ and splitting it into upper ({U}) and lower ({L})\ triangular parts ${\rm \textbf{t}_{\rm U}}$\ and ${\rm \textbf{t}_{\rm L}}$, respectively, the general sequence of transformations is given by \cite{malm86,olse95} 
\begin{equation}\label{eq:xtoa-general}
\varphi^{\rm X} = \varphi^{\rm A} {\rm \textbf{t}_{U}} + \varphi^{\rm X} {\rm \textbf{t}_{L}}\ .
\end{equation}
From the last equation, it follows that $\rm{\textbf{C}^{\rm XA}}$\ can be written as \cite{malm86,olse95} 
\begin{equation}\label{eq:xa-LU}
\rm{\textbf{C}^{\rm XA}} = \left({\rm \textbf{1} - \textbf{t}_{L}}\right) {\rm \textbf{t}_{U}}^{-1}\ ,
\end{equation}
or, alternatively, in terms of an LU-factorization \cite{malm86,olse95}\ as
\begin{equation}\label{eq:LU-mpstransmat}
{\rm \bf C}^{\rm XA} = {\rm \textbf{LU}}\ ,
\end{equation}
with $ {\rm \textbf{L}}$\ being a lower triangular matrix with unit diagonal elements and $ {\rm \textbf{U}}$\ an upper triangular matrix. 
Combining Eqs.~(\ref{eq:xa-LU})\ and (\ref{eq:LU-mpstransmat})\ we can determine the upper and lower triangular parts of \textbf{t} 
\begin{equation}\label{eq:t-parts}
{\rm \textbf{t}_{U}} = {{\rm \textbf{U}}}^{-1} \qquad \mbox{and}\qquad {\rm \textbf{t}_{L}} = {\rm \textbf{1} - {\rm \textbf{L}}}
\end{equation}
The same considerations hold for ${\rm \bf C}^{\rm YB}$\ and its factorization into upper and lower triangular matrices.

\paragraph{MPS wave function transformation}\label{par:mps-trafo}

With the elements of the matrix \textbf{t} at hand (\textit{vide supra}), the wave function expansion parameters can be transformed by a sequence 
of single-orbital transformations \cite{malm86,olse95}. A detailed account of this transformation approach can be found for CI-type wave functions in Refs. \citenum{malm86} and \citenum{olse95}. Following closely their \textit{ansatz}, we outline below the essential steps required to transform an MPS-type wave function representation from an expansion in the original basis $\{\varphi^{\rm X}_{p}\}$\ to an expansion in a biorthonormal basis $\{\varphi^{\rm A}_{p}\}$. As discussed in Subsection \ref{subsec:expMPSMPOnonoo}, the latter will allow us to calculate $N$-particle transition density matrices in a framework of standard second-quantization algebra. 

We start by transforming the wave function $\left|\Psi^{\rm }\right>$
\begin{equation}
\left|\Psi^{\rm}\right> = \sum\limits_{\boldsymbol{k}^X} M^{k_1^{\rm X}} M^{k_2^{\rm X}} \cdots M^{k_L^{\rm X}} \left|\boldsymbol{k}^X\right>\ ,
\end{equation}
where we have introduced above an additional superscript $\rm X$\ to emphasize that the MPS tensors of $\ket{\Psi^{\rm }}$\ 
refer to the original orbital basis $\{\varphi^{\rm X}\}$. With the parameters ${\rm \textbf{t}_{\rm L}}$\ 
and ${\rm \textbf{t}_{\rm U}}$\ calculated from an LU-factorization of ${\rm \bf C}^{\rm XA}$ according to Eq.~(\ref{eq:t-parts}), 
the MPS wave function transformation proceeds differently for inactive and active orbitals. 
No special action is required for the secondary (virtual) orbital space.

\subparagraph{Transformation with respect to the \textit{\underline{inactive}} orbital space}

As shown in Ref.~\citenum{malm86}, this transformation step reduces for CAS-type wave functions to a simple scaling of $\left|{\Psi^{\rm}}\right>$\ 
by a factor $\alpha$,
	\begin{eqnarray}\label{eq:trafo-p1}
	\left|{\Psi^{}}\right> & \equiv & \alpha \cdot \left|{\Psi^{}}\right> = \sum\limits_{\boldsymbol{k}^{\rm X}} \alpha \cdot \left(M^{k_1^{\rm X}} M^{k_2^{\rm X}} \cdots M^{k_L^{\rm X}} \right) \left|\boldsymbol{k}^{\rm X}\right>
	\end{eqnarray}
	with 
	\begin{equation}\label{eq:trafo-p2}
	\alpha = \prod_{i=1}^{\rm n_{I}} (t_{ii})^2\ ,
	\end{equation}
where the inactive orbital space comprises $\rm n_{I}$\ orbitals. 

\subparagraph{Transformation with respect to the \textit{\underline{active}} orbital space}
Assuming an active orbital space comprising $L$\ active orbitals, we set the orbital counter to $j = 1$\ and proceed as follows: 
	\begin{enumerate}
	\item Scale the $j$-th MPS tensor $M^{k_j^{\rm X^{}}}$\ of $\ket{\Psi^{\rm }}$\ consisting of a set of matrices (one for each basis state occupation $k_j^{\rm X}$) 
	with respect to the occupation number of the $j$-th orbital,
	\begin{equation}
	M^{k_j^{\rm X^{}}} \equiv 
	\left\{ 
	\begin{array}{r@{\hspace{6pt}}l}
	t_{jj}^2 \cdot M^{k_{j,\left|\uparrow\downarrow\right>}^{\rm X^{}}} & \mbox{for}\ k_j^{\rm X} = \left|\uparrow\downarrow\right> \\
	t_{jj} \cdot M^{k_{j,\left|\uparrow\right>}^{\rm X^{}}} & \mbox{for}\ k_j^{\rm X} = \left|\uparrow\right> \\
	t_{jj} \cdot M^{k_{j,\left|\downarrow\right>}^{\rm X^{}}} & \mbox{for}\ k_j^{\rm X} = \left|\downarrow\right> \\
	M^{k_{j,\left|0\right>}^{\rm X^{}}} & \mbox{for}\ k_j^{\rm X} = \left|0\right> \\
	\end{array}
	\right. \ ,
	\end{equation}
	and 
	\begin{equation}
	 M^{k_i^{\rm X^{}}} \equiv M^{k_i^{\rm X^{}}} \quad \forall\ i \neq j \land\ i = 1,\ldots,L\ .
	\end{equation}
	\item Construct a new MPO $\widehat{\mathcal{W}}$\ for the one-electron operator $\widehat{\mathcal{T}}$,
	\begin{equation}
	\widehat{\mathcal{T}} = \sum\limits_{m \neq j\ \sigma}^{L} \frac{t_{mj}}{t_{jj}}\  a_{m \sigma}^{\dagger} a_{j \sigma}\ ,
	\end{equation} 
	where the coefficients $t_{mj}$\ are given by the matrix elements of ${\rm \textbf{t}_{L}}$\ and ${\rm \textbf{t}_{U}}$, respectively. 
	Note that $\sigma$\ denotes here the eigenvalue of the spin eigenfunction. 
	\item Carry out a sequence of MPO--MPS operations \cite{scho11}, starting with the \textit{exact} product (analogous to Section 5.1, page 140$f$\ of Ref.~\citenum{scho11}) 
\begin{eqnarray}\label{eq:tx-1}
\ket{\Psi^{{(1)}}} & = & \widehat{\mathcal{W}}\ket{\Psi^{{{}}}} \nonumber \\
& = & \sum\limits_{\boldsymbol{k}^{\rm X} \boldsymbol{k^{\rm X \prime}}} \left(W^{k_1^{\rm X} k_1^{\rm X \prime}} W^{k_2^{\rm X} k_2^{\rm X \prime}} \cdots W^{k_L^{\rm X} k_L^{\rm X \prime}}\right) \left(M^{k^{\rm X \prime}_1} M^{k^{\rm X \prime}_2} \cdots M^{k^{\rm X \prime}_L}\right) \ket{\boldsymbol{k}^{\rm X}} \nonumber\\
& = & \sum\limits_{\boldsymbol{k}^{\rm X} \boldsymbol{k^{\rm X \prime}}} \sum\limits_{\boldsymbol{a} \boldsymbol{b}} \left(W^{k_1^{\rm X} k_1^{\rm X \prime}}_{1 b_1} W^{k_2^{\rm X} k_2^{\rm X \prime}}_{b_1 b_2} \cdots W^{k_L^{\rm X} k_L^{\rm X \prime}}_{b_{L-1} 1}\right) \left(M^{k^{\rm X \prime}_1}_{1 a_1} M^{k^{\rm X \prime}_2}_{a_1 a_2} \cdots M^{k^{\rm X \prime}_L}_{a_{L-1} 1}\right) \ket{\boldsymbol{k}^{\rm X}} \nonumber\\
& = & \sum\limits_{\boldsymbol{k}^{\rm X} \boldsymbol{k^{\rm X \prime}}} \sum\limits_{\boldsymbol{a} \boldsymbol{b}} \left(W^{k_1^{\rm X} k_1^{\rm X \prime}}_{1 b_1} M^{k^{\rm X \prime}_1}_{1 a_1}\right) \left(W^{k_2^{\rm X} k_2^{\rm X \prime}}_{b_1 b_2} M^{k^{\rm X \prime}_2}_{a_1 a_2}\right) \cdots \left(W^{k_L^{\rm X} k_L^{\rm X \prime}}_{b_{L-1} 1} M^{k^{\rm X \prime}_L}_{a_{L-1} 1}\right) \ket{\boldsymbol{k}^{\rm X}} \nonumber\\
& = & \sum\limits_{\boldsymbol{k}^{\rm X}} \sum\limits_{\boldsymbol{a} \boldsymbol{b}} N^{k_1^{\rm X}}_{\left(1 1\right) \left(b_1 a_1\right)}  
N^{k_2^{\rm X}}_{\left(b_1 a_1\right) \left(b_2 a_2\right)}  \cdots N^{k_L^{\rm X}}_{\left(b_{L-1} a_{L-1}\right) \left(1 1\right)}  \ket{\boldsymbol{k}^{\rm X}} \nonumber \\
& = & \sum\limits_{\boldsymbol{k}^{\rm X}} N^{k_1^{\rm X}}   N^{k_2^{\rm X}} \cdots N^{k_L^{\rm X}} \ket{\boldsymbol{k}^{\rm X}}\ ,
\end{eqnarray}
where the compact notation 
\begin{equation}
N^{k_i^{\rm X}}_{\left(b_{i-1} a_{i-1}\right) \left(b_i a_i\right)} = \sum_{k_i^{\rm X \prime}} W^{k_i^{\rm X} k_i^{\rm X \prime}}_{b_{i-1} b_i} M^{k^{\rm X \prime}_i}_{a_{i-1} a_i}\ ,
\end{equation}
has been introduced in the second to last step on the right-hand side of Eq.~(\ref{eq:tx-1}).
\item Compress $\left|{\Psi^{{(1)}}}\right>$\ by means of a singular-value decomposition (SVD) with a truncation threshold $\epsilon$\ of (at most) $\epsilon = 10^{-8}$. This is followed by
\begin{eqnarray}    
\ket{\Psi^{{(2)}}} &= &\widehat{\mathcal{W}}\ket{\Psi^{{(1)}}}\ \nonumber\\
& = & \sum\limits_{\boldsymbol{k}^{\rm X} \boldsymbol{k^{\rm X \prime}}} \left(W^{k_1^{\rm X} k_1^{\rm X \prime}} W^{k_2^{\rm X} k_2^{\rm X \prime}} \cdots W^{k_L^{\rm X} k_L^{\rm X \prime}}\right) \left(N^{k_1^{\rm X}}   N^{k_2^{\rm X}} \cdots N^{k_L^{\rm X}}\right) \ket{\boldsymbol{k}^{\rm X}} \nonumber\\ 
& = & \sum\limits_{\boldsymbol{k}^{\rm X}} \sum\limits_{\boldsymbol{a} \boldsymbol{b}} O^{k_1^{\rm X}}_{\left(1 1\right) \left(b_1 a_1\right)}  
O^{k_2^{\rm X}}_{\left(b_1 a_1\right) \left(b_2 a_2\right)}  \cdots O^{k_L^{\rm X}}_{\left(b_{L-1} a_{L-1}\right) \left(1 1\right)}  \ket{\boldsymbol{k}^{\rm X}} \nonumber \\
& = & \sum\limits_{\boldsymbol{k}^{\rm X}} O^{k_1^{\rm X}}   O^{k_2^{\rm X}} \cdots O^{k_L^{\rm X}} \ket{\boldsymbol{k}^{\rm X}}\ ,
\end{eqnarray}
in complete analogy to Eq.~(\ref{eq:tx-1}).
      \item Set $\left|{\Psi^{{}}}\right> \equiv \left|{\Psi^{{}}}\right> + \left|{\Psi^{{(1)}}}\right> + \frac{1}{2}\left|{\Psi^{{(2)}}}\right>$\ and 
      perform an SVD compression of $\left|{\Psi^{{}}}\right>$.
      \item Increase the orbital counter $j$\ by one.  If $j \leq L$ repeat steps (1)-(5), otherwise exit the transformation algorithm. Note that the 
      final MPS tensors of $\left|{\Psi^{{}}}\right>$\ now refer to an expansion in the biorthonormal basis $\{\bar{\varphi}^{\rm A}\}$, for example, 
      \begin{equation}
       M^{k_i^{\rm A^{}}} \equiv M^{k_i^{\rm X^{}}} \quad \forall\ i = 1,\ldots,L\ ,
      \end{equation}
      such that we can express $\left|{\Psi^{{}}}\right>$\ as 
     \begin{equation}
	\left|{\Psi^{}}\right> = \sum\limits_{\boldsymbol{k}^{\rm A}} M^{k_1^{\rm A}} M^{k_2^{\rm A}} \cdots M^{k_L^{\rm A}}\left|\boldsymbol{k}^{\rm A}\right>
     \end{equation}   
\end{enumerate}
      
Having found a representation of $\ket{\Psi}$\ in the biorthonormal basis $\{\bar{\varphi}^{\rm A}_{p}\}$, 
we repeat the transformation steps above for the inactive and active orbital spaces of the MPS 
wave function $\ket{\Phi^{}}$\ with ${\rm \textbf{t}_{\rm L}}$\ and ${\rm \textbf{t}_{\rm U}}$\ calculated from an 
LU-factorization of ${\rm \bf C}^{\rm YB}$.

\section{The State-Interaction Approach for MPS Wave Functions}\label{sec:si-mps}

\begin{figure*}[t]
	\centering
	\includegraphics[width=1.1\textwidth]{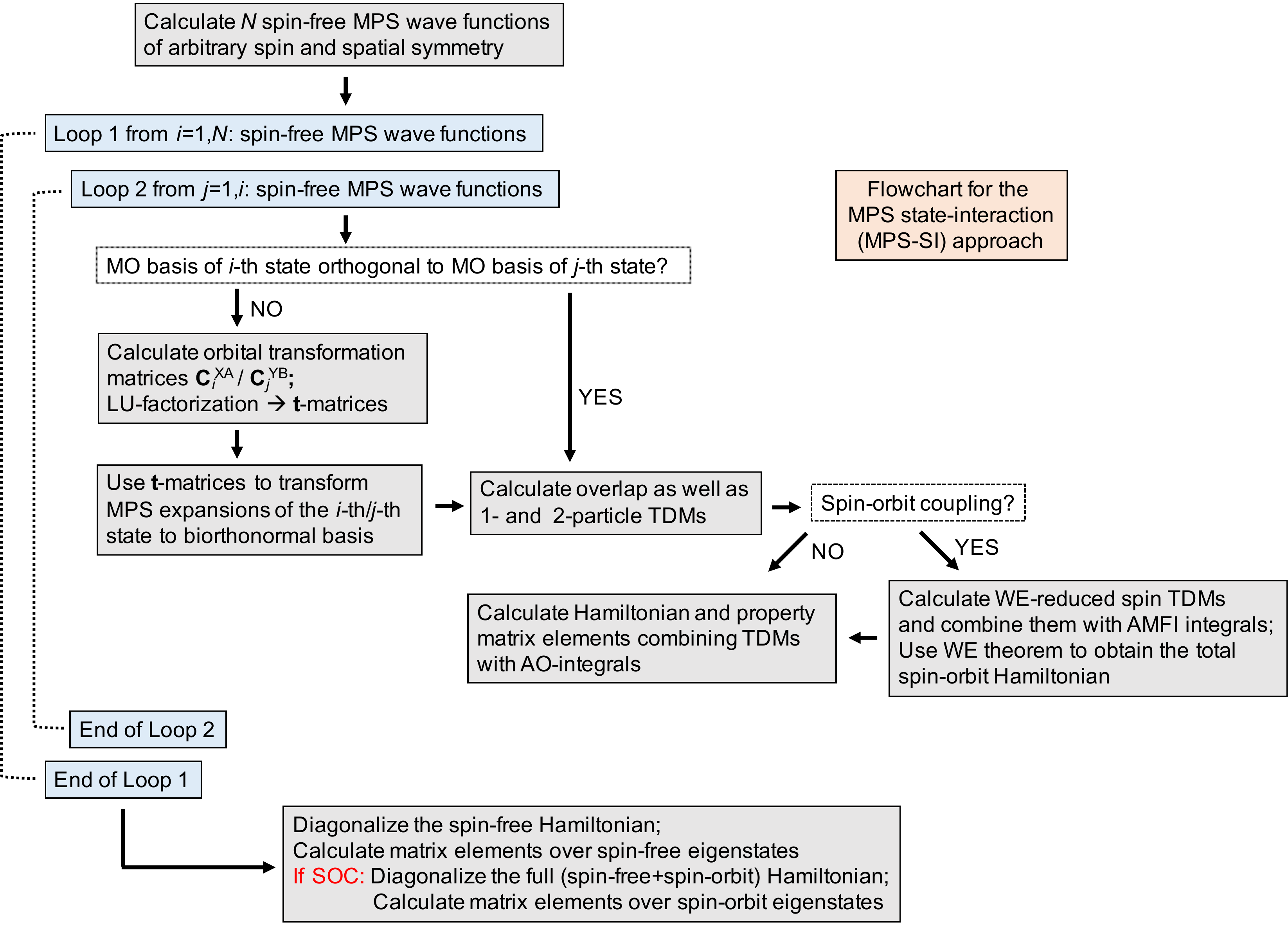}
	\caption{The flowchart illustrates the MPS wave function state-interaction (MPS-SI) approach targeting property matrix elements of 
		spin-free and spin-orbit coupled wave functions.}
	\label{fig:mpssi-flowchart}
\end{figure*}

With the overlap and transition matrix elements at hand, calculated
from spin-free MPS wave functions (i.e., SU(2) invariant MPS wave
functions $\ket{\Psi(S)}$\ with a well-defined total spin quantum
number $S$ as, for instance implemented in the \textsc{QCMaquis}
program \cite{kell16}) in a biorthonormal basis, we are now able not
only to calculate Hamiltonian but also matrix elements for arbitrary
one- and two-particle property operators such as transition dipole
moments (from which we can calculate oscillator strengths), angular
momentum eigenvalues and magnetic transition dipoles, or electric
field gradients.  Based on the discussion in the previous section,
Fig.~\ref{fig:mpssi-flowchart}\ illustrates the typical workflow of
our MPS-SI approach where the MPSs are obtained from different
DMRG-SCF optimizations.  We implemented this scheme in a development
version of the \textsc{Molcas} 8.0 \cite{molc15}\ software which, to a
large extent, exploits the existing framework of the CASSI/RASSI
implementation for CI-type wave functions by Malmqvist and co-workers
\cite{malm89,malm02}. Further details on the SI approach common to
both CI-type and MPS wave functions can be found in
Refs.~\citenum{malm89}\ and \citenum{malm02}\ and we here provide only
a brief summary of the most important steps as outlined in
Fig.~\ref{fig:mpssi-flowchart}.

Considering only the lower triangular matrix of dimension $N(N+1)$\
originating from $N$\ converged spin-free (non- or
scalar-relativistic) MPS wave functions of arbitrary spatial symmetry
and total spin $S$, we first test whether the $i$-th MPS shares the
same MO basis as the $j$-th state.  If the outcome of the test is negative, both MPSs
are transformed to the biorthonormal basis according to Section
\ref{par:mps-trafo}, before proceeding with the calculation of the wave
function overlap and transition density matrix elements.  The latter
are then combined with the corresponding AO-integrals to evaluate
property and Hamiltonian matrix elements. In the last step we
diagonalize the Hamiltonian matrix and calculate property matrix
elements over the resulting $N$ spin-free eigenstates.

Moreover, the calculation of magnetic properties such as g-tensors and hyperfine coupling tensors 
as well as the prediction of intersystem crossing rates requires access to SO coupled wave functions. 
To this end, we diagonalize the full Hamiltonian 
\begin{equation}
\label{eq:fullH}
\hat{\mathcal{H}} = \hat{\mathcal{H}}^{\rm el,sf} + \hat{\mathcal{H}}^{\rm SO}\ ,
\end{equation}
represented in the basis of the multiplet of states $\ket{\Psi(S,M)}$\ that can be obtained from a spin-free calculation of $\ket{\Psi(S)}$. 
Here, the state label $(S,M)$\ indicates the pair of total spin $S$\ and spin magnetic quantum number $M$\ 
the latter which can take values in the range from $-S$\ to $S$\ with unit increments. 
In our current approach, $\hat{\mathcal{H}}^{\rm el,sf}$ is typically a scalar-relativistic, spin-free electronic Hamiltonian, 
for example the Douglas-Kroll Hess Hamiltonian \cite{doug74,hess86,reih06} or the 
exact two-component Hamiltonian (X2C) \cite{jens05,kutz05,saue11,peng12} with the energy expectation 
value $E_{\rm A}$ for the eigenfunction $\ket{\Psi^{\rm A}(S,M)}$\ of state A. 
For $\hat{\mathcal{H}}^{\rm SO}$\ we consider an effective one-electron mean-field SO 
Hamiltonian \cite{hess96}\ in which the two-electron terms essentially serve as screening corrections 
of the (dominating) one-electron terms. 
Without significant loss of accuracy \cite{tatc99,nees05} we apply an atomic-mean field approximation 
of the SO integrals as implemented in the AMFI program of B.\ Schimmelpfennig \cite{amfi96} where all multicenter integrals are neglected. 
In the AMFI representation the effective one-electron mean-field SO Hamiltonian reads \cite{malm02}
\begin{equation}
\label{eq:HSO}
\hat{\mathcal{H}}^{\rm SO} = \sum\limits_{pq} \left(V_{pq}^{x} \hat{\rm T}_{pq}^{x} + V_{pq}^{y} \hat{\rm T}_{pq}^{y} + V_{pq}^{z} \hat{\rm T}_{pq}^{z}\right)\ ,
\end{equation}
where \textbf{V}$_{pq}$\ are the AO spin-orbit interaction integrals and \textbf{T}$_{pq}$\ the triplet operators in Cartesian representation. 
The latter are related to the more common triplet operators in spin-tensor form \cite{helg00}   
\begin{align}
{\hat{\rm {{T}}}}^{1,1}_{pq} &= -{a}^{\dagger}_{p\alpha} {a}^{}_{q\beta}\ ,\\
{\hat{\rm {{T}}}}^{1,0}_{pq} &= \frac{1}{\sqrt{2}} \left({a}^{\dagger}_{p\alpha} {a}^{}_{q\alpha} - {a}^{\dagger}_{p\beta} {a}^{}_{q\beta}\right)\ ,\\
{\hat{\rm {{T}}}}^{1,-1}_{pq} &= {a}^{\dagger}_{p\beta} {a}^{}_{q\alpha}\ ,
\end{align}
through a linear transformation \cite{helg00},
\begin{equation}
\left(\hat{\rm T}_{pq}^{x},\hat{\rm T}_{pq}^{y},\hat{\rm T}_{pq}^{z}\right) = \left({\hat{\rm {{T}}}}^{1,1}_{pq},{\hat{\rm {{T}}}}^{1,-1}_{pq},{\hat{\rm {{T}}}}^{1,0}_{pq}\right)
\left(
\begin{array}{rrr}
-\frac{1}{2}& -\frac{1}{2\imath}& 0 \\
\frac{1}{2}& -\frac{1}{2\imath}& 0 \\
0 & 0 & \frac{1}{\sqrt{2}}
\end{array}
\right)\ .
\end{equation}
Here, ${\hat{\rm {{T}}}}^{S,M}$\ is a spin-tensor operator whose eigenfunctions are a tensor state with spin eigenvalues $S$\ and $M$. 
It is one of the $2S+1$\ spin-tensor operators comprising the spin-tensor operator ${\hat{\rm {\textbf{T}}}}^{S}$\ of half-integer or integer rank $S$. 
To calculate the matrix representation of the SO operator given in Eq.~(\ref{eq:HSO})\ 
we need to evaluate matrix elements 
between state $\ket{\Psi^{\rm A}(S^{\prime},M^{\prime})}$ and state $\ket{\Psi^{\rm B}(S^{\prime\prime},M^{\prime\prime})}$ of the 
form $\left<\Psi^{\rm A}(S^{\prime},M^{\prime})\left|{\hat{\rm {{T}}}}^{S,M}\right|\Psi^{\rm B}(S^{\prime\prime},M^{\prime\prime})\right>$. 
These can be calculated very efficiently by employing the Wigner-Eckart (WE) theorem \cite{rose95,mari12}\ which 
states that the $(2S^{\prime}+1)\times(2S+1)\times(2S^{\prime\prime}+1)$ matrix elements above can be obtained as a product 
from a single (WE-reduced) matrix element $\left<\Psi^{\rm A}(S^{\prime})\left|\left|{\hat{\rm {\textbf{T}}}}^{S}\right|\right|\Psi^{\rm B}(S^{\prime\prime})\right>$\ and a Wigner 3$j$-symbol (calculated from Clebsch-Gordon coefficients),
\begin{eqnarray}
\left<\Psi^{\rm A}(S^{\prime},M^{\prime})\left|{\hat{\rm {{T}}}}^{S,M}\right|\Psi^{\rm B}(S^{\prime\prime},M^{\prime\prime})\right> & = &
(-1)^{S^{\prime\prime}+M^{\prime}-S} \nonumber \\
& & \times\left(
\begin{array}{rrr}
S^{\prime\prime}  & S & S^{\prime}\\
M^{\prime\prime} & M & M^{\prime}
\end{array}
\right)
\left<\Psi^{\rm A}(S^{\prime})\left|\left|{\hat{\rm {\textbf{T}}}}^{S}\right|\right|\Psi^{\rm B}(S^{\prime\prime})\right>\ .
\end{eqnarray}
Taking advantage of the WE theorem, we calculate WE-reduced one-particle (spin) transition density matrices \cite{malm02}
\begin{equation}
\Gamma^{\rm WE}_{pq} = \left<\Psi^{\rm A}(S^{\prime})\left|\left|{\hat{\rm {\textbf{T}}}}^{S}_{pq}\right|\right|\Psi^{\rm B}(S^{\prime\prime})\right>\ ,
\end{equation}
which can now be employed to calculate a WE-reduced matrix element \cite{malm02}
\begin{equation}
{\rm \textbf{V}^{\rm AB}} = \sum\limits_{pq} \Gamma^{\rm WE}_{pq} {\rm \textbf{V}}_{pq}\ .
\end{equation}
Since the matrix elements over spin states will only be non-zero if $\left|S^{\prime\prime} - S^{\prime}\right|$ is an integer $\le 1$,  
while also $\left|M^{\prime\prime} - M^{\prime}\right|$ is an integer $\le 1$, we are left with a total of nine non-zero SO 
Hamiltonian matrix elements, for example, 
\begin{equation}
\left<\Psi^{\rm A}(S^{}M^{})\left|\hat{H}^{\rm SO}\right|\Psi^{\rm B}(S+1^{}M\pm{1}^{})\right> = -\frac{\sqrt{\left(S{\pm}M+1\right)\left(S{\pm}M+2\right)}}{2}\left({\pm}V^{{\rm AB},x} + \imath V^{{\rm AB},y}\right)
\end{equation}
and
\begin{equation}
\left<\Psi^{\rm A}(S^{}M^{})\left|\hat{\mathcal{H}}^{\rm SO}\right|\Psi^{\rm B}(S^{}M^{})\right> = M V^{{\rm AB},z}\ ,
\end{equation}
where the remaining six cases can be found in Ref.~\citenum{malm02}.  

With the matrix elements of the SO operator at hand, we next diagonalize the total Hamiltonian matrix \textbf{H} of the Hamiltonian $\mathcal{H}$\ 
given in Eq.~(\ref{eq:fullH}) with the matrix elements given by 
\begin{equation}
\left<\Psi^{\rm A}(S^{\prime}M^{\prime})\left|\hat{\mathcal{H}}\right|\Psi^{\rm B}(S^{\prime\prime}M^{\prime\prime})\right> = {S}_{\rm AB} \delta_{S^{\prime}S^{\prime\prime}} \delta_{M^{\prime}M^{\prime\prime}} E_{\rm A} + 
\left<\Psi^{\rm A}(S^{\prime}M^{\prime})\left|\hat{\mathcal{H}}^{\rm SO}\right|\Psi^{\rm B}(S^{\prime\prime}M^{\prime\prime})\right>\ ,
\end{equation}
where ${S}_{\rm AB}$\ is the overlap between states A and B. Note that it is also possible, prior to the diagonalziation of \textbf{H},   
to correct the state energies for dynamic electron correlation contributions from, e.g., a DMRG-CASPT2 \cite{kura11,kura14,wout16b} 
or DMRG-NEVPT2 \cite{shar14,soko16,roem16,guos16,frei16} calculation 
by shifting the diagonal elements of the total Hamiltonian by \cite{malm02}
\begin{equation}
{\Delta}H_{\rm AB} = 0.5\left({\Delta}E_{\rm A} + {\Delta}E_{\rm B}\right) {S}_{\rm AB}\ ,
\end{equation}
where ${\Delta}E_{\rm A}$\ is the dynamical electron correlation contribution shift for state A. 

Diagonalization of \textbf{H} yields a set of SO coupled eigenstates which are linear combinations of all MPS wave functions with different spin and magnetic quantum numbers and the corresponding eigenvalues. In the final step (lower box in Fig.~\ref{fig:mpssi-flowchart}), we can now calculate the desired property matrix elements such as g-tensor matrix elements (see for example Ref.~\citenum{bolv06} for further details) in the basis of the SO eigenstates. 

\section{Numerical examples}\label{sec:secNUMEX}

\subsection{Closedness of MPS wave function expansion}\label{sec:closedMPS}

The transformation algorithm outlined in Section \ref{par:mps-trafo} is based on Malmqvist's approach \cite{malm86} which 
can only be applied to wave function expansions that are closed under de-excitation \cite{olse88,roos92,roos08a}. 
The latter implies that in each sequence of the wave function transformation no states outside of the 
original wave function expansion are generated. This prerequisite is fulfilled by full CI wave 
functions and was also shown to be the case for restricted active space wave functions \cite{olse88}.    
Here, we probe the closedness and (approximate) orbital rotation invariance of our MPS wave function transformation 
by systematically increasing the numerical accuracy of the MPS with a varying number of renormalized block states $m$ for a 
given active orbital space. As an example, we choose to study the zero-field splitting (ZFS)\ of the spin-free 
ground state $^3\Sigma_g^{-}$\ of the tellurium dimer Te$_2$\ for which CASSCF-SO reference data are available \cite{rota11} for 
both the ZFS\ and the SO matrix element $V$\ between the spin-free $^3\Sigma_g^-$\ and $^1\Sigma_g^+$\ states. 

In accordance with the reference work of Rota \textit{et al.} \cite{rota11}, 
we employ an all-electron ANO-RCC basis set of TZP quality \cite{anoM04, anoA05} for Te 
along with the 2nd-order Douglas-Kroll-Hess (DKH2) scalar-relativistic Hamiltonian \cite{wolf02}. 
The internuclear Te-Te distance of 2.557 \AA\ is taken from experiment \cite{hube79}. In the chalcogenide dimers, 
the low-lying electronic spectrum originates from the valence $(\pi^{\ast})^2$\ configuration which gives rise to 
the $X\ 0_g^{+}$\ electronic ground and $X_2\ 1_g, a\ 2_g$\ and $b\ 0g^{+}$\ excited states. We consider a CAS(8,6) with 
8 electrons allowed to freely occupy the six spatial orbitals ($\sigma, \sigma^{\ast}, \pi$\ and $\pi^{\ast}$) resulting from combinations of 
the atomic Te $5p$\ valence orbitals. 
For this active orbital space, we carried out individual state-averaged spin-free DMRG-SCF calculations for five triplet and six singlet states  
which are subsequently allowed to mix through SO interaction in our MPS-SI\ \textit{ansatz}. The latter requires a transformation to a biorthonormal 
MO basis and a corresponding rotation of the MPS wave functions to this basis. We therefore expect that any significant violation of the closedness 
and/or orbital rotation invariance of the MPS \textit{ansatz} for finite values of $m$\ would lead to considerable differences in  
the calculated spectrum and/or the SO matrix element V compared to the reference CASSCF data.  

All results are compiled in Table \ref{tab:0}. We first note that for the small CAS(8,6)\ space all our DMRG-SCF-SO data for the ZFS and the SO matrix element $V$, 
irrespective of the number of renormalized block states $m$\ ranging from $m=64$\ to $m=2048$, is in excellent agreement with the reference 
CASSCF-SO data (third-to-last row in Table \ref{tab:0}) obtained in this work. This result may not be surprising at first glance because of the size 
of the considered CAS but it nevertheless indicates that the closedness assumption holds for full CI-type wave functions represented by 
MPS expansions with finite $m$\ values. Moreover, orbital rotation invariance which is in principle only fullfilled exactly for 
MPS wave function expansions with $m \rightarrow \infty$\ appear not to have an impact on our MPS transformation approach 
to the biorthonormal basis. 

Compared to the CASSCF-SO data by Rota \textit{et al.}, our DMRG-SCF-SO/CASSCF-SO data deviate 
by several hundred wave numbers for the excitation energies of the spin-free and SO coupled excited 
states as well as by 25 cm$^{-1}$ for the SO matrix element $V$. 
This is partly due to the fact that we took into account only the lowest five (six) triplet (singlet) states in the 
wave function optimization. 
Considering all spin-free states up to 70000 cm$^{-1}$\ (second-to-last row in Table \ref{tab:0})\ partly reduces the 
observed deviations, in particular for $V$. In summary, we emphasize that our CASSCF-SO and DMRG-SCF-SO data 
are consistent. 

\begin{table}[h!]
	\caption{\label{tab:0} Relative first excited state energies of Te$_2$\ as obtained from spin-free (SF) and spin-orbit coupled (SOC) DMRG-SCF 
		calculations with a varying number of renormalized block states $m$. $V$\ denotes the spin-orbit coupling matrix element 
		between the $^3\Sigma_g^-$\ and $^1\Sigma_g^+$ states. All values are given in cm$^{-1}$.}
				\begin{tabular}{l@{\hspace{4mm}}l@{\hspace{4mm}}l@{\hspace{4mm}}ll@{\hspace{4mm}}lll@{\hspace{5mm}}c}\hline\hline
					& & & \multicolumn{2}{c}{SF} & \multicolumn{3}{c}{SOC} & \\ \cline{4-5} \cline{6-8}
			method	& CAS & $m$ & $^1{\Delta}_{g}$ & $^1\Sigma_g^+$ & $X_2$1$_g$ & $a$2$_g$ & $b$ $0_{g}^{+}$ & $V$\\
			DMRG-SCF-SO& (8,6) & 64  & 3712 & 6588 & 1759 & 5471 & 10107 & 3832\\
			DMRG-SCF-SO& (8,6) & 128  & 3711 & 6587 & 1759 & 5471 & 10106 & 3832\\
			DMRG-SCF-SO& (8,6) & 256  & 3711 & 6587 & 1759 & 5470 & 10105 & 3832\\
			DMRG-SCF-SO& (8,6) & 512  & 3711 & 6586 & 1759 & 5470 & 10105 & 3832\\
			DMRG-SCF-SO& (8,6) & 1024 & 3711 & 6587 & 1759 & 5470 & 10105 & 3832\\
			DMRG-SCF-SO& (8,6) & 2048 & 3711 & 6586 & 1759 & 5470 & 10105 & 3832\\
			CASSCF-SO$^a$  & (8,6) & - & 3711 & 6586 & 1759 & 5470 & 10105 & 3832\\
			CASSCF-SO$^b$  & (8,6) & - & 3507 & 6298 & 1657 & 4917 & 9317 & 3808\\
			CASSCF-SO$^c$ & (8,6) & - & 2716 & 6009 & 1700 & 4180 & 9153 & 3807\\ \hline
			    \multicolumn{9}{l}{$^a$ This work.}\\
			     \multicolumn{9}{l}{$^b$ This work. All spin-free states up to 70,000 cm$^{-1}$\ were considered.}\\
				\multicolumn{9}{l}{$^c$ Ref.~\citenum{rota11}; CAS(8,6)SCF-SO/TZP data.}
				\end{tabular}
\end{table}

\subsection{Magnetic resonance properties of sample $f^1$\ and $f^2$\ actinide complexes}\label{sec:magprop}

To illustrate the capabilities of our MPS-SI approach, 
we calculate g-factors of prototypical $f^1$- and $f^2$-type actinide 
complexes, namely NpO$_2^{2+}$\ and  PuO$_2^{2+}$, and compare 
to CASSCF/CASPT2 data from Gendron \textit{et al.} \cite{gend14}. 
To facilitate a comparison, we follow the same computational protocol as 
outlined in Ref.~\citenum{gend14}\ which we briefly summarize here. 
We employed all-electron ANO-RCC basis sets of TZP quality \cite{anoM04, anoA05} 
along with the scalar-relativistic DKH2 Hamiltonian \cite{wolf02} and a Cholesky-decomposition (CD) of the 
two-electron repulsion integrals as implemented in \textsc{Molcas} \cite{molc15} (keyword \texttt{Cholesky} in the integral module \textsc{SEWARD}). 
Dynamical electron correlation is included through CD-DMRG-NEVPT2 (second order \textit{n}-electron valence perturbation theory) 
calculations \cite{knec16,frei16}\ with a state-averaged DMRG-SCF reference wave function. 
Active orbital spaces for the latter comprise 7 (8) electrons in 10 orbitals for NpO$_2^{2+}$ (PuO$_2^{2+}$) denoted in the following as CAS(7,10) (CAS(8,10)). For NpO$_2^{2+}$, we simultaneously optimized six doublet states while for PuO$_2^{2+}$\ we carried out individual state-averaged 
DMRG-SCF calculations for 20 triplet and 20 singlet states, respectively. 
In all DMRG-SCF calculations, the number renormalized block states $m$ was set to 1024 which is sufficient to 
yield results of CASSCF quality for the active orbital spaces considered. In accordance with Gendron \textit{et al.} \cite{gend14}, C$_1$\ point group symmetry 
was assumed in calculations.

As outlined in Section \ref{sec:si-mps}, our MPS-SI  
approach is integrated in the CASSI/RASSI framework \cite{malm89,malm02}\ of \textsc{Molcas}\ which allowed us
 to calculate the SO operator matrix elements from the contraction of WE-reduced spin transition 
 density matrices with atomic mean-field integrals \cite{hess96,amfi96}\ while EPR g-factors were calculated based 
on the approach of Bolvin \cite{bolv06}.

In the case of Np(VI)O$_2^{2+}$\ we carried out additional 
four-component EPR calculations using a multi-reference CI (MRCI) approach \cite{vad13} based on 
the Dirac-Coulomb Hamiltonian and all-electron 
uncontracted cc-pVTZ \cite{dunn89} and dyall.v3z \cite{dyal07b} basis sets for 
oxygen and neptunium, respectively. MP2 natural spinors 
obtained from correlating the Np 5s5p5d6s6p and O 2s2p 
electrons while keeping the remaining core electrons of Np and O frozen 
constitute the orbital basis for the MRCI expansion. 
In the MP2 calculation, a virtual spinor threshold was set to 20 hartree 
such that the virtual correlation space comprised all 
recommended core-valence and valence-correlation functions. 
The MRCI active space includes all occupied spinors with occupation numbers less than 1.98 as well as all virtual spinors 
with occupation numbers up to about 0.001. 
The active space was further split in three subspaces with a maximum 
allowed excitation level of singles and singles-doubles from the first space containing 
all spinors with occupation larger than $1.80$\ into a model space 
comprising the four partially occupied nonbonding Np $5f$\ spinors of Np(VI)O$_2^{2+}$ \cite{gend14}. 
From the combined two lower spaces a maximum of 
singles-doubles-triples excitations are allowed 
into the third space comprising in total 90 spinors. 

Table \ref{tab:1}\ summarizes our property calculations for 
NpO$_2^{2+}$\ and PuO$_2^{2+}$. 
Starting with NpO$_2^{2+}$, we note that our  
DMRG/NEVPT2-SO results are in close agreement with the corresponding 
CASSCF/CASPT2-SO data from Gendron \textit{et al.} \cite{gend14}\ both for the g-factors,  g$_{\parallel}$ and g$_{\bot}$, respectively, 
and the vertical excitation energy $\Delta E$\ from the $^2\Phi_{5/2u}$\ 
ground to the $^2\Delta_{3/2u}$\ first excited state. 
The DMRG/NEVPT2-SO g$_{\parallel}=4.235$ value for NpO$_2^{2+}$\ compares further well with our 
reference four-component MRCI g$_{\parallel}=4.283$ value which indicates that the considered number of spin-free states is 
sufficient to achieve convergence of the SO coupling contributions. 
Moreover, we find that the ground- and excited state compositions of 
the SO coupled  $^2\Phi_{5/2u}$\ and $^2\Delta_{3/2u}$\ states in terms of their $\phi, \delta$\ and $\pi$\ character 
are well reproduced in our DMRG/NEVPT2-SO calculation in comparison to the CASSCF/CASPT2-SO reference data, 
in particular the sizable admixture of $12\%\ \delta$ character in the SO coupled $^2\Phi_{5/2u}$\ ground state 
which originates from a strong coupling of the spin-free $^2\Phi_{u}$\ and $^2\Delta_{u}$\ states \cite{gend14}.

Turning next to PuO$_2^{2+}$, we find an excellent agreement of our DMRG-SCF-SO results for 
the electronic $\Omega = 4_g$\ ground and the lowest-lying excited states with the corresponding CASSCF-SO 
reference data by Gendron \textit{et al.} \cite{gend14}. We are not only able to correctly reproduce the ground-state 
g-factors of g$_{\parallel}=6.076$ and g$_{\bot}=0.000$\ but also find matching vertical excitation energies $\Delta E$\ with 
deviations of at most 5 cm$^{-1}$\ for the first three excited states. Similar to NpO$_2^{2+}$, the lowest spin-free 
states of the plutonyl ion originate from occupations of the non-bonding $\delta$\ and $\phi$\ orbitals\ with different 
combinations (parallel or antiparallel) of spin ($M_S$)\ and angular momentum ($M_L$) projections. 
In accordance with Hund’s rules, the spin-free ground state is a $^3H_g$\ ($\delta^1 \phi^1$\ occupation) triplet state with parallel 
angular momentum projections ($M_L = \pm 2 \pm 3 = \pm 5$) and antiparallel spin projection M$_S = \mp 1$. 
Further combinations of $M_S$\ and $M_L$\ within the $\delta^1 \phi^1$\ occupation manifold lead to $^3\Sigma^{-}_{g}$\ and $^3\Pi_g$\ states 
which we find at 3875 cm$^{-1}$\ and 6563 cm$^{-1}$\ above the $^3H_g$\ ground state, respectively.
Moreover, according to our DMRG-SCF calculations the first spin-singlet state $^1\Sigma^{+}_g$\ is located at 
10384 cm$^{-1}$\ above the electronic ground state.

At the SO level, the $^3H_g$\ term splits into its three components with $\Omega = 4_g, 5_g$, and $6_g$\ which are then 
allowed to mix with states of the same $\Omega$. Although the $\Omega = 4_g$\ ground state remains predominantly of $^3H_g$\ 
character (96\%), it exhibits an admixture of the $^1\Gamma_g$\ state (2\%)\ which is energetically separated from the $^3H_g$\ state 
by about 15300 cm$^{-1}$\ in the spin-free framework. In addition, as can be seen from Table \ref{tab:1}, the first two SO-coupled excited states 
of $\Omega = 0^+_g$\ and $\Omega = 1_g$\ at $\Delta E = 4383$\ cm$^{-1}$\ and 6985 cm$^{-1}$, respectively, originate from a strong 
admixture of different triplet and singlet states. Their state composition, as obtained from our DMRG-SCF-SO calculations,  
compares nicely with the CASSCF-SO data of Gendron \textit{et al.} \cite{gend14}.    

\begin{table}[h!]
\caption{\label{tab:1} Relative energies $\Delta E$ [cm$^{-1}$], g-factors (absolute values), and state composition 
of the two lowest-energy electronic states of NpO$_2^{2+}$\ as well as the 
lowest electronic states of PuO$_2^{2+}$ as obtained from 
DMRG-SCF-SO and DMRG/NEVPT2-SO calculations. 
The same active orbital space was employed as for the CASSCF-SO/CASPT2-SO data 
taken from the work of Gendron \textit{et al.} \cite{gend14}. For details on the 4c-MRCI calculation, see text.}
				\begin{tabular}{ll@{\hspace{4mm}}llll}\hline\hline
                    \multicolumn{6}{c}{NpO$_2^{2+}$}\\
   & & DMRG-SCF-SO & NEVPT2-SO & CASPT2-SO$^a$ & 4c-MRCI\\
ground state    & $\Delta E$       & 0 & 0 & 0 & 0 \\
$^2\Phi_{5/2u}$ & g$_{\parallel}$ & 4.228 & 4.235 & 4.233 & 4.283 \\
                & g$_{\bot}^{b}$      & 0.002 & 0.002  & 0.002  & 0.000 \\
                & composition     & 89\% $\phi$, 11\% $\delta$ & 88\% $\phi$, 12\% $\delta$ & 88\% $\phi$, 12\% $\delta$  & - \\ \hline
1st excited state & $\Delta E$       & 3294 & 3086 & 3107 & - \\
$^2\Delta_{3/2u}$ & g$_{\parallel}$ & 2.025 & 2.035 & 2.037 & - \\
                  & g$_{\bot}^{b}$      & 0.002 & 0.017 & 0.005 & - \\
                  & composition     & 99\% $\delta$, 1\% $\pi$ & 98\% $\delta$, 2\% $\pi$ & 98\% $\delta$, 2\% $\pi$  & - \\ \hline
					& & & & & \\ \hline
                    \multicolumn{6}{c}{PuO$_2^{2+}$}\\
   & & \multicolumn{2}{c}{DMRG-SCF-SO} & CASSCF-SO$^a$ & \\
ground state    & $\Delta E$         &  \multicolumn{2}{c}{0}     & 0 & \\
$4_g$               & g$_{\parallel}$  & \multicolumn{2}{c}{6.076} & 6.076 & \\
                          & g$_{\bot}$        & \multicolumn{2}{c}{0.000} & 0.000 & \\
                         & composition      & \multicolumn{2}{c}{96\% $^3H_g$, 2\% $^1\Gamma_g$} & 96\% $^3H_g$ & \\ \hline
1st excited state     & $\Delta E$     &  \multicolumn{2}{c}{4383}     & 4388 & \\
$0_g^{+}$               & composition       &  \multicolumn{2}{c}{57\% $^3\Sigma_g^{-}$, 27\% $^3\Pi_g$, 14\% $^1\Sigma_g^{+}$} & 57\% $^3\Sigma_g^{-}$, 27\% $^3\Pi_g$, 14\% $^1\Sigma_g^{+}$& \\ \hline
2nd excited state     & $\Delta E$     &  \multicolumn{2}{c}{6985}     & 6985 & \\
$1_g$               & composition       &  \multicolumn{2}{c}{36\% $^3\Sigma_g^{-}$, 48\% $^3\Pi_g$, 14\% $^1\Pi_g$ } & 35\% $^3\Sigma_g^{-}$, 50\% $^3\Pi_g$, 14\% $^1\Pi_g$ & \\ \hline
3rd excited state     & $\Delta E$     &  \multicolumn{2}{c}{7150}     & 7152 & \\
$5_g$               & composition       &  \multicolumn{2}{c}{99\% $^3H_g$} & 99\% $^3H_g$ & \\ \hline                                               
					\multicolumn{6}{l}{$^a$ Ref.~\citenum{gend14}; CAS(7,10)PT2-SO data for NpO$_2^{2+}$\ and
CAS(8,10)SCF-SO data for PuO$_2^{2+}$.}\\
\multicolumn{6}{l}{$^b$ A non-zero value for g$_{\bot}$\ indicates a minor symmetry-breaking in the optimized spin-free wave}\\
\multicolumn{6}{l}{functions due to the neglect of point group symmetry.}
				\end{tabular}
\end{table}

\section{Conclusions and Outlook}\label{sec:conclusions}

In this work, we presented a state-interaction (SI) approach for nonorthogonal matrix product state (MPS) 
wave functions dubbed as MPS-SI where the MPSs can be obtained from several different (state-specific) DMRG-SCF 
wave function optimizations. Our MPS-SI approach complements existing 
implementations for traditional CI-type wave functions while generalizing earlier SI approaches for DMRG wave 
functions to nonorthogonal wave functions as basis set for the MPS-SI approach.

Allowing for nonorthogonality of the input MPS wave functions adds a greater flexibility to the preceeding orbital optimization 
step, which will be particularly beneficial for transition metal and lanthanide/actinide compounds, 
where state-specific optimizations allow us to explicitly consider different open-shell $d$- or $f$-orbital 
occupations. Moreover, our nonorthogonal MPS-SI approach no longer requires 
a state averaging of molecular states with different spin symmetry at the orbital optimization 
step if, for example, spin-orbit coupling elements between these states are to be calculated. 
In this work, we discussed the transformation of orbitals and MPS wave functions from a nonorthogonal 
to a biorthonormal basis by formulating a nonunitary orbital transformation algorithm for MPS wave functions. 
This opens up for an efficient calculation of overlap matrix elements between nonorthogonal MPS wave functions 
and matrix elements of the one- and two-particle reduced transition density matrix as well as of the spin-orbit operator.  
After diagonalization of the resulting spin-free and spin-orbit Hamiltonian matrices target properties are calculated 
in the basis of the corresponding \textit{orthogonal and non-interacting} (spin-orbit coupled) eigenstates. 

We demonstrated the applicability of the nonorthogonal MPS-SI approach 
for two example open-shell $f^{n}$ actinide molecules, namely 
NpO$_2^{2+}$ ($f^{1}$\ complex) and PuO$_2^{2+}$ ($f^2$\ complex) 
for which we calculated ground- and excited-state properties including 
g-factors. Our spin-orbit DMRG-SCF/NEVPT2 g-factors for the two lowest doublet states of NpO$_2^{2+}$\ agree well with corresponding CASSCF/CASPT2 data reported in the literature. The ground-state g-factors are close to the values obtained from a four-component 
MRCI g-tensor calculation which includes spin-orbit coupling variationally from the outset. 
This indicates that the property values are (close to being) converged with respect to the number of spin-free states included in our 
MPS-SI approach. 
Subsequently, we carried out DMRG-SCF g-factor calculations for PuO$_2^{2+}$\ by 
allowing the lowest spin-free singlet and triplet states to interact through spin-orbit coupling. 
We found a similar agreement of our DMRG-SCF results 
to the CASSCF values as was the case for NpO$_2^{2+}$\ which illustrates 
that our nonorthogonal MPS-SI approach works equally well for cases where the 
interacting states do not share the same molecular orbital basis. 
Combined with our recently developed Cholesky-decomposition DMRG-NEVPT2 implementation, 
our MPS-SI approach will be valuable not only for the study of (magnetic) properties of transition metal and/or heavy-element 
complexes in different spin states but also for the full exploration of the photophysics 
and, more generally, excited-state (surface hopping) 
dynamics of chromophores and light-harvesting materials. The latter 
requires, besides the calculation of non-adiabatic coupling elements 
to locate and characterize conical intersections, the determination 
of intersystem crossing probabilities between electronic states of 
different spin multiplicity induced by spin-orbit coupling.

\section*{Acknowledgments}
This work was supported by the Schweizer Nationalfonds. SK is grateful to Prof. H. J. Aa. Jensen (SDU Odense) for providing him access to the four-component MRCI-EPR module. JA acknowledges support from the U.S.\ Department of Energy,
Office of Basic Energy Sciences, Heavy Element Chemistry program,
under grant DE-SC0001136 (formerly DE-FG02-09ER16066). 

\renewcommand{\theequation}{A\arabic{equation}}
\setcounter{equation}{0}  

\section*{Appendix}

\subsection*{The action of an annihilation operator in a nonorthogonal basis}

An annihilation operator in a nonorthogonal spin-orbital basis acting on 
an ONV is considered here. 
We follow closely the derivation that can be found in Ref.~\citenum{helg00}. Assume an ONV for $L$\ spin-orbitals 
as it was given in Eq.~(\ref{eq:onvDEF}). 
Recalling that annihilation of the vacuum state is impossible, 
\begin{equation}\label{app:eq:2}
a_{q\sigma} \ket{\rm {vac}} = \sum_{p}^{} {S}^{1/2}_{qp}  \underset{= 0\ {\rm \forall p}}{\underbrace{{\tilde{a}^{}_{p\sigma}} \ket{\rm {vac}}}} = 0\ ,
\end{equation}
we write the action of an annihilation operator on the ONV as (anti-)commutator of the annihilation operator and the product of even (odd) numbers of creation operators
\begin{eqnarray}\label{app:eq:3}
a_{q\sigma} \ket{\boldsymbol{k}_{}} & = & \left(\vphantom{\underset{= 0}{\underbrace{a_{q\sigma} \left(\ \prod_{p=1}^{L} \left(a^{\dagger}_{p\sigma} \right)^{k_{p\sigma}}\right) }}} a_{q\sigma} \left(\ \prod_{p=1}^{L} \left(a^{\dagger}_{p\sigma} \right)^{k_{p\sigma}}\right) \right. \left. \mp \underset{= 0}{\underbrace{\left(\ \prod_{p=1}^{L} \left(a^{\dagger}_{p\sigma} \right)^{k_{p\sigma}}\right) a_{q\sigma}}}\right) \ket{\rm {vac}} \nonumber\\
& = & 
\left\{ 
\begin{array}{l@{\hspace{8pt}}l}
\left[a_{q\sigma}, \prod\limits_{p=1}^{L} \left(a^{\dagger}_{p\sigma} \right)^{k_{p\sigma}} \right] \ket{\rm {vac}} & {\rm for\ even}\ L\\ 
& \\
\left\{a_{q\sigma}, \prod\limits_{p=1}^{L} \left(a^{\dagger}_{p\sigma} \right)^{k_{p\sigma}} \right\} \ket{\rm {vac}} & {\rm for\ odd}\ L
\end{array}
\right.\ .
\end{eqnarray}
The (anti-)commutator appearing in Eq.~(\ref{app:eq:3})\ can be written in terms of linear combinations of a basic anticommutator 
of an annihilation and an creation operator exploiting the corresponding operator identity
\begin{equation}\label{app:eq:4}
\left.
\begin{array}{l}
\left[\mathcal{\hat{A}}, \mathcal{\hat{B}}_1 \cdots \mathcal{\hat{B}}_L\right] \\
\\
\left\{\mathcal{\hat{A}}, \mathcal{\hat{B}}_1 \cdots \mathcal{\hat{B}}_L\right\} 
 \end{array}
 \right\} = \sum_{p=1}^{L} (-1)^{p-1} \mathcal{\hat{B}}_1 \cdots \left\{\mathcal{\hat{A}}, \mathcal{\hat{B}}_p\right\} \cdots \mathcal{\hat{B}}_L\ .
\end{equation}
Inserting Eq.~(\ref{app:eq:4})\ into the right-hand side of Eq.~(\ref{app:eq:3})\ yields 
\begin{eqnarray}\label{app:eq:5}
a_{q\sigma} \ket{\boldsymbol{k}} &=&  \sum_{p=1}^{L} k_{p\sigma} (-1)^{\left[\sum\limits_{j=1}^{p-1} k_{j\sigma}\right]} 
\left(a^{\dagger}_{1\sigma}\right)^{k_{1\sigma}}\cdots \underset{= S_{qp} \delta_{\sigma\sigma}}{\underbrace{\left\{a_{q\sigma}, \left(a_{p\sigma}^{\dagger}\right)^{k_{p\sigma}}\right\}}}\cdots\left(a_{L\sigma}^{\dagger}\right)^{k_{L\sigma}}
\ket{\rm {vac}}\ .
\end{eqnarray}
Resolving the anticommutator by the identity in Eq.~(\ref{eq:anticomm_CREANNnoo}), as indicated in Eq.~(\ref{app:eq:5}), yields the final result
\begin{equation}\label{app:eq:6}
 a_{q\sigma} \ket{\boldsymbol{k}_{}} =  \sum_{p=1}^{L} k_{p\sigma} (-1)^{\left[\sum\limits_{j=1}^{p-1} k_{j\sigma}\right]} S_{qp} \ket{k_{1} k_{2} \ldots 0_{p} \ldots k_{L}}\ .
\end{equation}


\begin{thebibliography}{109}%
\makeatletter
\providecommand \@ifxundefined [1]{%
 \@ifx{#1\undefined}
}%
\providecommand \@ifnum [1]{%
 \ifnum #1\expandafter \@firstoftwo
 \else \expandafter \@secondoftwo
 \fi
}%
\providecommand \@ifx [1]{%
 \ifx #1\expandafter \@firstoftwo
 \else \expandafter \@secondoftwo
 \fi
}%
\providecommand \natexlab [1]{#1}%
\providecommand \enquote  [1]{``#1''}%
\providecommand \bibnamefont  [1]{#1}%
\providecommand \bibfnamefont [1]{#1}%
\providecommand \citenamefont [1]{#1}%
\providecommand \href@noop [0]{\@secondoftwo}%
\providecommand \href [0]{\begingroup \@sanitize@url \@href}%
\providecommand \@href[1]{\@@startlink{#1}\@@href}%
\providecommand \@@href[1]{\endgroup#1\@@endlink}%
\providecommand \@sanitize@url [0]{\catcode `\\12\catcode `\$12\catcode
  `\&12\catcode `\#12\catcode `\^12\catcode `\_12\catcode `\%12\relax}%
\providecommand \@@startlink[1]{}%
\providecommand \@@endlink[0]{}%
\providecommand \url  [0]{\begingroup\@sanitize@url \@url }%
\providecommand \@url [1]{\endgroup\@href {#1}{\urlprefix }}%
\providecommand \urlprefix  [0]{URL }%
\providecommand \Eprint [0]{\href }%
\providecommand \doibase [0]{http://dx.doi.org/}%
\providecommand \selectlanguage [0]{\@gobble}%
\providecommand \bibinfo  [0]{\@secondoftwo}%
\providecommand \bibfield  [0]{\@secondoftwo}%
\providecommand \translation [1]{[#1]}%
\providecommand \BibitemOpen [0]{}%
\providecommand \bibitemStop [0]{}%
\providecommand \bibitemNoStop [0]{.\EOS\space}%
\providecommand \EOS [0]{\spacefactor3000\relax}%
\providecommand \BibitemShut  [1]{\csname bibitem#1\endcsname}%
\let\auto@bib@innerbib\@empty
\bibitem [{\citenamefont {Barbatti}(2011)}]{barb11}%
  \BibitemOpen
  \bibfield  {author} {\bibinfo {author} {\bibfnamefont {M.}~\bibnamefont
  {Barbatti}},\ }\href@noop {} {\bibfield  {journal} {\bibinfo  {journal}
  {{WIRE}s Comput. Mol. Sci.}\ }\textbf {\bibinfo {volume} {1}},\ \bibinfo
  {pages} {620} (\bibinfo {year} {2011})}\BibitemShut {NoStop}%
\bibitem [{\citenamefont {Mai}, \citenamefont {Marquetand},\ and\ \citenamefont
  {Gonz{\'a}lez}(2015)}]{mai15}%
  \BibitemOpen
  \bibfield  {author} {\bibinfo {author} {\bibfnamefont {S.}~\bibnamefont
  {Mai}}, \bibinfo {author} {\bibfnamefont {P.}~\bibnamefont {Marquetand}}, \
  and\ \bibinfo {author} {\bibfnamefont {L.}~\bibnamefont {Gonz{\'a}lez}},\
  }\href@noop {} {\bibfield  {journal} {\bibinfo  {journal} {Int. J. Quant.
  Chem.}\ }\textbf {\bibinfo {volume} {115}},\ \bibinfo {pages} {1215}
  (\bibinfo {year} {2015})}\BibitemShut {NoStop}%
\bibitem [{\citenamefont {Suaud}\ \emph {et~al.}(2009)\citenamefont {Suaud},
  \citenamefont {Bonnet}, \citenamefont {Boilleau}, \citenamefont
  {Lab{\`e}guerie},\ and\ \citenamefont {Guih{\'e}ry}}]{suau09}%
  \BibitemOpen
  \bibfield  {author} {\bibinfo {author} {\bibfnamefont {N.}~\bibnamefont
  {Suaud}}, \bibinfo {author} {\bibfnamefont {M.-L.}\ \bibnamefont {Bonnet}},
  \bibinfo {author} {\bibfnamefont {C.}~\bibnamefont {Boilleau}}, \bibinfo
  {author} {\bibfnamefont {P.}~\bibnamefont {Lab{\`e}guerie}}, \ and\ \bibinfo
  {author} {\bibfnamefont {N.}~\bibnamefont {Guih{\'e}ry}},\ }\href@noop {}
  {\bibfield  {journal} {\bibinfo  {journal} {J. Am. Chem. Soc.}\ }\textbf
  {\bibinfo {volume} {131}},\ \bibinfo {pages} {715} (\bibinfo {year}
  {2009})}\BibitemShut {NoStop}%
\bibitem [{\citenamefont {Chang}, \citenamefont {Fedro},\ and\ \citenamefont
  {van Veenendaal}(2010)}]{chan10}%
  \BibitemOpen
  \bibfield  {author} {\bibinfo {author} {\bibfnamefont {J.}~\bibnamefont
  {Chang}}, \bibinfo {author} {\bibfnamefont {A.~J.}\ \bibnamefont {Fedro}}, \
  and\ \bibinfo {author} {\bibfnamefont {M.}~\bibnamefont {van Veenendaal}},\
  }\href@noop {} {\bibfield  {journal} {\bibinfo  {journal} {Phys. Rev. B}\
  }\textbf {\bibinfo {volume} {82}},\ \bibinfo {pages} {075124} (\bibinfo
  {year} {2010})}\BibitemShut {NoStop}%
\bibitem [{\citenamefont {Marian}(2012)}]{mari12}%
  \BibitemOpen
  \bibfield  {author} {\bibinfo {author} {\bibfnamefont {C.~M.}\ \bibnamefont
  {Marian}},\ }\href@noop {} {\bibfield  {journal} {\bibinfo  {journal}
  {{WIREs} Comp. Mol. Sci.}\ }\textbf {\bibinfo {volume} {2}},\ \bibinfo
  {pages} {187} (\bibinfo {year} {2012})}\BibitemShut {NoStop}%
\bibitem [{\citenamefont {Kaupp}, \citenamefont {B{\"u}hl},\ and\ \citenamefont
  {Malkin}(2004)}]{kaup04}%
  \BibitemOpen
  \bibinfo {editor} {\bibfnamefont {M.}~\bibnamefont {Kaupp}}, \bibinfo
  {editor} {\bibfnamefont {M.}~\bibnamefont {B{\"u}hl}}, \ and\ \bibinfo
  {editor} {\bibfnamefont {V.~G.}\ \bibnamefont {Malkin}},\ eds.,\ \href@noop
  {} {\emph {\bibinfo {title} {Calculation of {NMR} and {EPR} Parameters:
  Theory and Applications}}}\ (\bibinfo  {publisher} {Wiley-VCH},\ \bibinfo
  {year} {2004})\BibitemShut {NoStop}%
\bibitem [{\citenamefont {Gerloch}\ and\ \citenamefont
  {McMeeking}(1975)}]{gerl75}%
  \BibitemOpen
  \bibfield  {author} {\bibinfo {author} {\bibfnamefont {M.}~\bibnamefont
  {Gerloch}}\ and\ \bibinfo {author} {\bibfnamefont {R.~F.}\ \bibnamefont
  {McMeeking}},\ }\href@noop {} {\bibfield  {journal} {\bibinfo  {journal} {J.
  Chem. Soc., Dalton Trans.}\ ,\ \bibinfo {pages} {2443}} (\bibinfo {year}
  {1975})}\BibitemShut {NoStop}%
\bibitem [{\citenamefont {Bolvin}\ and\ \citenamefont
  {Autschbach}(2016)}]{Autschbach:2015a}%
  \BibitemOpen
  \bibfield  {author} {\bibinfo {author} {\bibfnamefont {H.}~\bibnamefont
  {Bolvin}}\ and\ \bibinfo {author} {\bibfnamefont {J.}~\bibnamefont
  {Autschbach}},\ }in\ \href {http://dx.doi.org/10.1007/978-3-642-41611-8_12-1}
  {\emph {\bibinfo {booktitle} {Handbook of Relativistic Quantum Chemistry}}},\
  \bibinfo {editor} {edited by\ \bibinfo {editor} {\bibfnamefont
  {W.}~\bibnamefont {Liu}}}\ (\bibinfo  {publisher} {Springer},\ \bibinfo
  {address} {Berlin},\ \bibinfo {year} {2016})\BibitemShut {NoStop}%
\bibitem [{\citenamefont {Bolvin}(2006)}]{bolv06}%
  \BibitemOpen
  \bibfield  {author} {\bibinfo {author} {\bibfnamefont {H.}~\bibnamefont
  {Bolvin}},\ }\href@noop {} {\bibfield  {journal} {\bibinfo  {journal}
  {ChemPhysChem}\ }\textbf {\bibinfo {volume} {7}},\ \bibinfo {pages} {1575}
  (\bibinfo {year} {2006})}\BibitemShut {NoStop}%
\bibitem [{\citenamefont {Ganyushin}\ and\ \citenamefont
  {Neese}(2013)}]{gany12}%
  \BibitemOpen
  \bibfield  {author} {\bibinfo {author} {\bibfnamefont {D.}~\bibnamefont
  {Ganyushin}}\ and\ \bibinfo {author} {\bibfnamefont {F.}~\bibnamefont
  {Neese}},\ }\href@noop {} {\bibfield  {journal} {\bibinfo  {journal} {J.
  Chem. Phys.}\ }\textbf {\bibinfo {volume} {138}},\ \bibinfo {pages} {104113}
  (\bibinfo {year} {2013})}\BibitemShut {NoStop}%
\bibitem [{\citenamefont {Vad}\ \emph {et~al.}(2013)\citenamefont {Vad},
  \citenamefont {Pedersen}, \citenamefont {N{\o}rager},\ and\ \citenamefont
  {Jensen}}]{vad13}%
  \BibitemOpen
  \bibfield  {author} {\bibinfo {author} {\bibfnamefont {M.~S.}\ \bibnamefont
  {Vad}}, \bibinfo {author} {\bibfnamefont {M.~N.}\ \bibnamefont {Pedersen}},
  \bibinfo {author} {\bibfnamefont {A.}~\bibnamefont {N{\o}rager}}, \ and\
  \bibinfo {author} {\bibfnamefont {H.~J.~A.}\ \bibnamefont {Jensen}},\
  }\href@noop {} {\bibfield  {journal} {\bibinfo  {journal} {J. Chem. Phys.}\
  }\textbf {\bibinfo {volume} {138}},\ \bibinfo {pages} {214106} (\bibinfo
  {year} {2013})}\BibitemShut {NoStop}%
\bibitem [{\citenamefont {Sharkas}, \citenamefont {Pritchard},\ and\
  \citenamefont {Autschbach}(2015)}]{Autschbach:2014o}%
  \BibitemOpen
  \bibfield  {author} {\bibinfo {author} {\bibfnamefont {K.}~\bibnamefont
  {Sharkas}}, \bibinfo {author} {\bibfnamefont {B.}~\bibnamefont {Pritchard}},
  \ and\ \bibinfo {author} {\bibfnamefont {J.}~\bibnamefont {Autschbach}},\
  }\href {http://dx.doi.org/10.1021/ct500988h} {\bibfield  {journal} {\bibinfo
  {journal} {J.\ Chem.\ Theory Comput.}\ }\textbf {\bibinfo {volume} {11}},\
  \bibinfo {pages} {538} (\bibinfo {year} {2015})}\BibitemShut {NoStop}%
\bibitem [{\citenamefont {van Lenthe}, \citenamefont {Wormer},\ and\
  \citenamefont {van~der Avoird}(1997)}]{vanl97}%
  \BibitemOpen
  \bibfield  {author} {\bibinfo {author} {\bibfnamefont {E.}~\bibnamefont {van
  Lenthe}}, \bibinfo {author} {\bibfnamefont {P.~E.~S.}\ \bibnamefont
  {Wormer}}, \ and\ \bibinfo {author} {\bibfnamefont {A.}~\bibnamefont {van~der
  Avoird}},\ }\href@noop {} {\bibfield  {journal} {\bibinfo  {journal} {J.
  Chem. Phys.}\ }\textbf {\bibinfo {volume} {107}},\ \bibinfo {pages} {2488}
  (\bibinfo {year} {1997})}\BibitemShut {NoStop}%
\bibitem [{\citenamefont {Repisky}\ \emph {et~al.}(2010)\citenamefont
  {Repisky}, \citenamefont {Komorovsky}, \citenamefont {Malkin}, \citenamefont
  {Malkina},\ and\ \citenamefont {Malkin}}]{repi10}%
  \BibitemOpen
  \bibfield  {author} {\bibinfo {author} {\bibfnamefont {M.}~\bibnamefont
  {Repisky}}, \bibinfo {author} {\bibfnamefont {S.}~\bibnamefont {Komorovsky}},
  \bibinfo {author} {\bibfnamefont {E.}~\bibnamefont {Malkin}}, \bibinfo
  {author} {\bibfnamefont {O.~L.}\ \bibnamefont {Malkina}}, \ and\ \bibinfo
  {author} {\bibfnamefont {V.~G.}\ \bibnamefont {Malkin}},\ }\href@noop {}
  {\bibfield  {journal} {\bibinfo  {journal} {Chem. Phys. Lett.}\ }\textbf
  {\bibinfo {volume} {488}},\ \bibinfo {pages} {94} (\bibinfo {year}
  {2010})}\BibitemShut {NoStop}%
\bibitem [{\citenamefont {Verma}\ and\ \citenamefont
  {Autschbach}(2013)}]{verm13}%
  \BibitemOpen
  \bibfield  {author} {\bibinfo {author} {\bibfnamefont {P.}~\bibnamefont
  {Verma}}\ and\ \bibinfo {author} {\bibfnamefont {J.}~\bibnamefont
  {Autschbach}},\ }\href@noop {} {\bibfield  {journal} {\bibinfo  {journal} {J.
  Chem. Theory Comp.}\ }\textbf {\bibinfo {volume} {9}},\ \bibinfo {pages}
  {1932} (\bibinfo {year} {2013})}\BibitemShut {NoStop}%
\bibitem [{\citenamefont {Lushington}\ and\ \citenamefont
  {Grein}(1996)}]{lush96}%
  \BibitemOpen
  \bibfield  {author} {\bibinfo {author} {\bibfnamefont {G.~H.}\ \bibnamefont
  {Lushington}}\ and\ \bibinfo {author} {\bibfnamefont {F.}~\bibnamefont
  {Grein}},\ }\href@noop {} {\bibfield  {journal} {\bibinfo  {journal} {Int. J.
  Quant. Chem.}\ }\textbf {\bibinfo {volume} {60}},\ \bibinfo {pages} {1679}
  (\bibinfo {year} {1996})}\BibitemShut {NoStop}%
\bibitem [{\citenamefont {Brownridge}\ \emph {et~al.}(2003)\citenamefont
  {Brownridge}, \citenamefont {Grein}, \citenamefont {Tatchen}, \citenamefont
  {Kleinschmidt},\ and\ \citenamefont {Marian}}]{brow03}%
  \BibitemOpen
  \bibfield  {author} {\bibinfo {author} {\bibfnamefont {S.}~\bibnamefont
  {Brownridge}}, \bibinfo {author} {\bibfnamefont {F.}~\bibnamefont {Grein}},
  \bibinfo {author} {\bibfnamefont {J.}~\bibnamefont {Tatchen}}, \bibinfo
  {author} {\bibfnamefont {M.}~\bibnamefont {Kleinschmidt}}, \ and\ \bibinfo
  {author} {\bibfnamefont {C.~M.}\ \bibnamefont {Marian}},\ }\href@noop {}
  {\bibfield  {journal} {\bibinfo  {journal} {J. Chem. Phys.}\ }\textbf
  {\bibinfo {volume} {118}},\ \bibinfo {pages} {9552} (\bibinfo {year}
  {2003})}\BibitemShut {NoStop}%
\bibitem [{\citenamefont {Neese}(2005)}]{nees05}%
  \BibitemOpen
  \bibfield  {author} {\bibinfo {author} {\bibfnamefont {F.}~\bibnamefont
  {Neese}},\ }\href@noop {} {\bibfield  {journal} {\bibinfo  {journal} {J.
  Chem. Phys.}\ }\textbf {\bibinfo {volume} {122}},\ \bibinfo {pages} {034107}
  (\bibinfo {year} {2005})}\BibitemShut {NoStop}%
\bibitem [{\citenamefont {Vancoillie}, \citenamefont {Malmqvist},\ and\
  \citenamefont {Pierloot}(2007)}]{vanc07}%
  \BibitemOpen
  \bibfield  {author} {\bibinfo {author} {\bibfnamefont {S.}~\bibnamefont
  {Vancoillie}}, \bibinfo {author} {\bibfnamefont {P.-{\AA}.}\ \bibnamefont
  {Malmqvist}}, \ and\ \bibinfo {author} {\bibfnamefont {K.}~\bibnamefont
  {Pierloot}},\ }\href@noop {} {\bibfield  {journal} {\bibinfo  {journal}
  {ChemPhysChem}\ }\textbf {\bibinfo {volume} {8}},\ \bibinfo {pages} {1803}
  (\bibinfo {year} {2007})}\BibitemShut {NoStop}%
\bibitem [{\citenamefont {Neese}(2007)}]{nees07a}%
  \BibitemOpen
  \bibfield  {author} {\bibinfo {author} {\bibfnamefont {F.}~\bibnamefont
  {Neese}},\ }\href@noop {} {\bibfield  {journal} {\bibinfo  {journal} {Mol.
  Phys.}\ }\textbf {\bibinfo {volume} {105}},\ \bibinfo {pages} {2507}
  (\bibinfo {year} {2007})}\BibitemShut {NoStop}%
\bibitem [{\citenamefont {Tatchen}, \citenamefont {Kleinschmidt},\ and\
  \citenamefont {Marian}(2009)}]{tatc09}%
  \BibitemOpen
  \bibfield  {author} {\bibinfo {author} {\bibfnamefont {J.}~\bibnamefont
  {Tatchen}}, \bibinfo {author} {\bibfnamefont {M.}~\bibnamefont
  {Kleinschmidt}}, \ and\ \bibinfo {author} {\bibfnamefont {C.~M.}\
  \bibnamefont {Marian}},\ }\href@noop {} {\bibfield  {journal} {\bibinfo
  {journal} {J. Chem. Phys.}\ }\textbf {\bibinfo {volume} {130}},\ \bibinfo
  {pages} {154106} (\bibinfo {year} {2009})}\BibitemShut {NoStop}%
\bibitem [{\citenamefont {Roemelt}(2015)}]{roem15}%
  \BibitemOpen
  \bibfield  {author} {\bibinfo {author} {\bibfnamefont {M.}~\bibnamefont
  {Roemelt}},\ }\href@noop {} {\bibfield  {journal} {\bibinfo  {journal} {J.
  Chem. Phys.}\ }\textbf {\bibinfo {volume} {143}},\ \bibinfo {pages} {044112}
  (\bibinfo {year} {2015})}\BibitemShut {NoStop}%
\bibitem [{\citenamefont {Sayfutyarova}\ and\ \citenamefont
  {Chan}(2016)}]{sayf16}%
  \BibitemOpen
  \bibfield  {author} {\bibinfo {author} {\bibfnamefont {E.~R.}\ \bibnamefont
  {Sayfutyarova}}\ and\ \bibinfo {author} {\bibfnamefont {G.~K.-L.}\
  \bibnamefont {Chan}},\ }\href@noop {} {\bibfield  {journal} {\bibinfo
  {journal} {J. Chem. Phys.}\ }\textbf {\bibinfo {volume} {144}},\ \bibinfo
  {pages} {234301} (\bibinfo {year} {2016})}\BibitemShut {NoStop}%
\bibitem [{\citenamefont {Szalay}\ \emph {et~al.}(2012)\citenamefont {Szalay},
  \citenamefont {M{\"u}ller}, \citenamefont {Gidofalvi}, \citenamefont
  {Lischka},\ and\ \citenamefont {Shepard}}]{szal12}%
  \BibitemOpen
  \bibfield  {author} {\bibinfo {author} {\bibfnamefont {P.~G.}\ \bibnamefont
  {Szalay}}, \bibinfo {author} {\bibfnamefont {T.}~\bibnamefont {M{\"u}ller}},
  \bibinfo {author} {\bibfnamefont {G.}~\bibnamefont {Gidofalvi}}, \bibinfo
  {author} {\bibfnamefont {H.}~\bibnamefont {Lischka}}, \ and\ \bibinfo
  {author} {\bibfnamefont {R.}~\bibnamefont {Shepard}},\ }\href@noop {}
  {\bibfield  {journal} {\bibinfo  {journal} {Chem. Rev.}\ }\textbf {\bibinfo
  {volume} {112}},\ \bibinfo {pages} {108} (\bibinfo {year}
  {2012})}\BibitemShut {NoStop}%
\bibitem [{\citenamefont {Roca-Sanjuan}, \citenamefont {Aquilante},\ and\
  \citenamefont {Lindh}(2012)}]{roca12}%
  \BibitemOpen
  \bibfield  {author} {\bibinfo {author} {\bibfnamefont {D.}~\bibnamefont
  {Roca-Sanjuan}}, \bibinfo {author} {\bibfnamefont {F.}~\bibnamefont
  {Aquilante}}, \ and\ \bibinfo {author} {\bibfnamefont {R.}~\bibnamefont
  {Lindh}},\ }\href@noop {} {\bibfield  {journal} {\bibinfo  {journal} {{WIREs}
  Comp. Mol. Sci.}\ }\textbf {\bibinfo {volume} {2}},\ \bibinfo {pages} {585}
  (\bibinfo {year} {2012})}\BibitemShut {NoStop}%
\bibitem [{\citenamefont {Stein}, \citenamefont {von Burg},\ and\ \citenamefont
  {Reiher}(2016)}]{stei16b}%
  \BibitemOpen
  \bibfield  {author} {\bibinfo {author} {\bibfnamefont {C.~J.}\ \bibnamefont
  {Stein}}, \bibinfo {author} {\bibfnamefont {V.}~\bibnamefont {von Burg}}, \
  and\ \bibinfo {author} {\bibfnamefont {M.}~\bibnamefont {Reiher}},\
  }\href@noop {} {\bibfield  {journal} {\bibinfo  {journal} {J. Chem. Theory
  Comput.}\ }\textbf {\bibinfo {volume} {12}},\ \bibinfo {pages} {3764}
  (\bibinfo {year} {2016})}\BibitemShut {NoStop}%
\bibitem [{\citenamefont {Olsen}(2011)}]{olse11}%
  \BibitemOpen
  \bibfield  {author} {\bibinfo {author} {\bibfnamefont {J.}~\bibnamefont
  {Olsen}},\ }\href@noop {} {\bibfield  {journal} {\bibinfo  {journal} {Int. J.
  Quantum Chem.}\ }\textbf {\bibinfo {volume} {111}},\ \bibinfo {pages} {3267}
  (\bibinfo {year} {2011})}\BibitemShut {NoStop}%
\bibitem [{\citenamefont {Stein}\ and\ \citenamefont {Reiher}(2016)}]{stei16a}%
  \BibitemOpen
  \bibfield  {author} {\bibinfo {author} {\bibfnamefont {C.~J.}\ \bibnamefont
  {Stein}}\ and\ \bibinfo {author} {\bibfnamefont {M.}~\bibnamefont {Reiher}},\
  }\href@noop {} {\bibfield  {journal} {\bibinfo  {journal} {J. Chem. Theory
  Comput.}\ }\textbf {\bibinfo {volume} {12}},\ \bibinfo {pages} {1760}
  (\bibinfo {year} {2016})}\BibitemShut {NoStop}%
\bibitem [{\citenamefont {Aquilante}\ \emph
  {et~al.}(2015{\natexlab{a}})\citenamefont {Aquilante}, \citenamefont
  {Autschbach}, \citenamefont {Carlson}, \citenamefont {Chibotaru},
  \citenamefont {Delcey}, \citenamefont {De~Vico}, \citenamefont
  {Fdez.~Galv{\'a}n}, \citenamefont {Ferr{\'e}}, \citenamefont {Frutos},
  \citenamefont {Gagliardi}, \citenamefont {Garavelli}, \citenamefont
  {Giussani}, \citenamefont {Hoyer}, \citenamefont {Li~Manni}, \citenamefont
  {Lischka}, \citenamefont {Ma}, \citenamefont {Malmqvist}, \citenamefont
  {M{\"u}ller}, \citenamefont {Nenov}, \citenamefont {Olivucci}, \citenamefont
  {Pedersen}, \citenamefont {Peng}, \citenamefont {Plasser}, \citenamefont
  {Pritchard}, \citenamefont {Reiher}, \citenamefont {Rivalta}, \citenamefont
  {Schapiro}, \citenamefont {Segarra-Mart{\'\i}}, \citenamefont {Stenrup},
  \citenamefont {Truhlar}, \citenamefont {Ungur}, \citenamefont {Valentini},
  \citenamefont {Vancoillie}, \citenamefont {Veryazov}, \citenamefont
  {Vysotskiy}, \citenamefont {Weingart}, \citenamefont {Zapata},\ and\
  \citenamefont {Lindh}}]{aqui15}%
  \BibitemOpen
  \bibfield  {author} {\bibinfo {author} {\bibfnamefont {F.}~\bibnamefont
  {Aquilante}}, \bibinfo {author} {\bibfnamefont {J.}~\bibnamefont
  {Autschbach}}, \bibinfo {author} {\bibfnamefont {R.~K.}\ \bibnamefont
  {Carlson}}, \bibinfo {author} {\bibfnamefont {L.~F.}\ \bibnamefont
  {Chibotaru}}, \bibinfo {author} {\bibfnamefont {M.~G.}\ \bibnamefont
  {Delcey}}, \bibinfo {author} {\bibfnamefont {L.}~\bibnamefont {De~Vico}},
  \bibinfo {author} {\bibfnamefont {I.}~\bibnamefont {Fdez.~Galv{\'a}n}},
  \bibinfo {author} {\bibfnamefont {N.}~\bibnamefont {Ferr{\'e}}}, \bibinfo
  {author} {\bibfnamefont {L.~M.}\ \bibnamefont {Frutos}}, \bibinfo {author}
  {\bibfnamefont {L.}~\bibnamefont {Gagliardi}}, \bibinfo {author}
  {\bibfnamefont {M.}~\bibnamefont {Garavelli}}, \bibinfo {author}
  {\bibfnamefont {A.}~\bibnamefont {Giussani}}, \bibinfo {author}
  {\bibfnamefont {C.~E.}\ \bibnamefont {Hoyer}}, \bibinfo {author}
  {\bibfnamefont {G.}~\bibnamefont {Li~Manni}}, \bibinfo {author}
  {\bibfnamefont {H.}~\bibnamefont {Lischka}}, \bibinfo {author} {\bibfnamefont
  {D.}~\bibnamefont {Ma}}, \bibinfo {author} {\bibfnamefont {P.~A.}\
  \bibnamefont {Malmqvist}}, \bibinfo {author} {\bibfnamefont {T.}~\bibnamefont
  {M{\"u}ller}}, \bibinfo {author} {\bibfnamefont {A.}~\bibnamefont {Nenov}},
  \bibinfo {author} {\bibfnamefont {M.}~\bibnamefont {Olivucci}}, \bibinfo
  {author} {\bibfnamefont {T.~B.}\ \bibnamefont {Pedersen}}, \bibinfo {author}
  {\bibfnamefont {D.}~\bibnamefont {Peng}}, \bibinfo {author} {\bibfnamefont
  {F.}~\bibnamefont {Plasser}}, \bibinfo {author} {\bibfnamefont
  {B.}~\bibnamefont {Pritchard}}, \bibinfo {author} {\bibfnamefont
  {M.}~\bibnamefont {Reiher}}, \bibinfo {author} {\bibfnamefont
  {I.}~\bibnamefont {Rivalta}}, \bibinfo {author} {\bibfnamefont
  {I.}~\bibnamefont {Schapiro}}, \bibinfo {author} {\bibfnamefont
  {J.}~\bibnamefont {Segarra-Mart{\'\i}}}, \bibinfo {author} {\bibfnamefont
  {M.}~\bibnamefont {Stenrup}}, \bibinfo {author} {\bibfnamefont {D.~G.}\
  \bibnamefont {Truhlar}}, \bibinfo {author} {\bibfnamefont {L.}~\bibnamefont
  {Ungur}}, \bibinfo {author} {\bibfnamefont {A.}~\bibnamefont {Valentini}},
  \bibinfo {author} {\bibfnamefont {S.}~\bibnamefont {Vancoillie}}, \bibinfo
  {author} {\bibfnamefont {V.}~\bibnamefont {Veryazov}}, \bibinfo {author}
  {\bibfnamefont {V.~P.}\ \bibnamefont {Vysotskiy}}, \bibinfo {author}
  {\bibfnamefont {O.}~\bibnamefont {Weingart}}, \bibinfo {author}
  {\bibfnamefont {F.}~\bibnamefont {Zapata}}, \ and\ \bibinfo {author}
  {\bibfnamefont {R.}~\bibnamefont {Lindh}},\ }\href@noop {} {\bibfield
  {journal} {\bibinfo  {journal} {J. Comput. Chem.}\ }\textbf {\bibinfo
  {volume} {37}},\ \bibinfo {pages} {506} (\bibinfo {year}
  {2015}{\natexlab{a}})}\BibitemShut {NoStop}%
\bibitem [{\citenamefont {White}(1992)}]{whit92}%
  \BibitemOpen
  \bibfield  {author} {\bibinfo {author} {\bibfnamefont {S.~R.}\ \bibnamefont
  {White}},\ }\href@noop {} {\bibfield  {journal} {\bibinfo  {journal} {Phys.
  Rev. Lett.}\ }\textbf {\bibinfo {volume} {69}},\ \bibinfo {pages} {2863}
  (\bibinfo {year} {1992})}\BibitemShut {NoStop}%
\bibitem [{\citenamefont {White}(1993)}]{whit93}%
  \BibitemOpen
  \bibfield  {author} {\bibinfo {author} {\bibfnamefont {S.~R.}\ \bibnamefont
  {White}},\ }\href@noop {} {\bibfield  {journal} {\bibinfo  {journal} {Phys.
  Rev. B}\ }\textbf {\bibinfo {volume} {48}},\ \bibinfo {pages} {10345}
  (\bibinfo {year} {1993})}\BibitemShut {NoStop}%
\bibitem [{\citenamefont {Schollw\"ock}(2005)}]{scho05}%
  \BibitemOpen
  \bibfield  {author} {\bibinfo {author} {\bibfnamefont {U.}~\bibnamefont
  {Schollw\"ock}},\ }\href@noop {} {\bibfield  {journal} {\bibinfo  {journal}
  {Rev.\ Mod.\ Phys.}\ }\textbf {\bibinfo {volume} {77}},\ \bibinfo {pages}
  {259} (\bibinfo {year} {2005})}\BibitemShut {NoStop}%
\bibitem [{\citenamefont {Schollw\"{o}ck}(2011)}]{scho11}%
  \BibitemOpen
  \bibfield  {author} {\bibinfo {author} {\bibfnamefont {U.}~\bibnamefont
  {Schollw\"{o}ck}},\ }\href@noop {} {\bibfield  {journal} {\bibinfo  {journal}
  {Ann. Phys.}\ }\textbf {\bibinfo {volume} {326}},\ \bibinfo {pages} {96}
  (\bibinfo {year} {2011})}\BibitemShut {NoStop}%
\bibitem [{\citenamefont {Legeza}\ \emph {et~al.}(2008)\citenamefont {Legeza},
  \citenamefont {Noack}, \citenamefont {Solyom},\ and\ \citenamefont
  {Tincani}}]{lege08}%
  \BibitemOpen
  \bibfield  {author} {\bibinfo {author} {\bibfnamefont {{\"O}.}~\bibnamefont
  {Legeza}}, \bibinfo {author} {\bibfnamefont {R.~M.}\ \bibnamefont {Noack}},
  \bibinfo {author} {\bibfnamefont {J.}~\bibnamefont {Solyom}}, \ and\ \bibinfo
  {author} {\bibfnamefont {L.}~\bibnamefont {Tincani}},\ }in\ \href@noop {}
  {\emph {\bibinfo {booktitle} {Computational Many-Particle Physics}}},\
  \bibinfo {series} {Lecture Notes in Physics}, Vol.\ \bibinfo {volume} {739},\
  \bibinfo {editor} {edited by\ \bibinfo {editor} {\bibfnamefont
  {H.}~\bibnamefont {Fehske}}, \bibinfo {editor} {\bibfnamefont
  {R.}~\bibnamefont {Schneider}}, \ and\ \bibinfo {editor} {\bibfnamefont
  {A.}~\bibnamefont {Weisse}}}\ (\bibinfo {year} {2008})\ pp.\ \bibinfo {pages}
  {653--664}\BibitemShut {NoStop}%
\bibitem [{\citenamefont {Chan}\ \emph {et~al.}(2008)\citenamefont {Chan},
  \citenamefont {Dorando}, \citenamefont {Ghosh}, \citenamefont {Hachmann},
  \citenamefont {Neuscamman}, \citenamefont {Wang},\ and\ \citenamefont
  {Yanai}}]{chan08}%
  \BibitemOpen
  \bibfield  {author} {\bibinfo {author} {\bibfnamefont {G.~K.-L.}\
  \bibnamefont {Chan}}, \bibinfo {author} {\bibfnamefont {J.~J.}\ \bibnamefont
  {Dorando}}, \bibinfo {author} {\bibfnamefont {D.}~\bibnamefont {Ghosh}},
  \bibinfo {author} {\bibfnamefont {J.}~\bibnamefont {Hachmann}}, \bibinfo
  {author} {\bibfnamefont {E.}~\bibnamefont {Neuscamman}}, \bibinfo {author}
  {\bibfnamefont {H.}~\bibnamefont {Wang}}, \ and\ \bibinfo {author}
  {\bibfnamefont {T.}~\bibnamefont {Yanai}},\ }in\ \href@noop {} {\emph
  {\bibinfo {booktitle} {Frontiers in {{Quantum Systems}} in {{Chemistry}} and
  {{Physics}}}}},\ \bibinfo {series and number} {Progress in Theoretical
  Chemistry and Physics},\ \bibinfo {editor} {edited by\ \bibinfo {editor}
  {\bibfnamefont {S.}~\bibnamefont {Wilson}}, \bibinfo {editor} {\bibfnamefont
  {P.~J.}\ \bibnamefont {Grout}}, \bibinfo {editor} {\bibfnamefont
  {J.}~\bibnamefont {Maruani}}, \bibinfo {editor} {\bibfnamefont
  {G.}~\bibnamefont {Delgado-Barrio}}, \ and\ \bibinfo {editor} {\bibfnamefont
  {P.}~\bibnamefont {Piecuch}}}\ (\bibinfo  {publisher} {{Springer
  Netherlands}},\ \bibinfo {year} {2008})\ pp.\ \bibinfo {pages}
  {49--65}\BibitemShut {NoStop}%
\bibitem [{\citenamefont {Marti}\ and\ \citenamefont {Reiher}(2010)}]{mart10}%
  \BibitemOpen
  \bibfield  {author} {\bibinfo {author} {\bibfnamefont {K.~H.}\ \bibnamefont
  {Marti}}\ and\ \bibinfo {author} {\bibfnamefont {M.}~\bibnamefont {Reiher}},\
  }\href@noop {} {\bibfield  {journal} {\bibinfo  {journal} {Z. Phys. Chem.}\
  }\textbf {\bibinfo {volume} {224}},\ \bibinfo {pages} {583} (\bibinfo {year}
  {2010})}\BibitemShut {NoStop}%
\bibitem [{\citenamefont {Marti}\ and\ \citenamefont {Reiher}(2011)}]{mart11}%
  \BibitemOpen
  \bibfield  {author} {\bibinfo {author} {\bibfnamefont {K.~H.}\ \bibnamefont
  {Marti}}\ and\ \bibinfo {author} {\bibfnamefont {M.}~\bibnamefont {Reiher}},\
  }\href@noop {} {\bibfield  {journal} {\bibinfo  {journal} {Phys. Chem. Chem.
  Phys.}\ }\textbf {\bibinfo {volume} {13}},\ \bibinfo {pages} {6750} (\bibinfo
  {year} {2011})}\BibitemShut {NoStop}%
\bibitem [{\citenamefont {Chan}\ and\ \citenamefont {Sharma}(2011)}]{chan11a}%
  \BibitemOpen
  \bibfield  {author} {\bibinfo {author} {\bibfnamefont {G.~K.-L.}\
  \bibnamefont {Chan}}\ and\ \bibinfo {author} {\bibfnamefont {S.}~\bibnamefont
  {Sharma}},\ }\href@noop {} {\bibfield  {journal} {\bibinfo  {journal} {Annu.
  Rev. Phys. Chem.}\ }\textbf {\bibinfo {volume} {62}},\ \bibinfo {pages} {465}
  (\bibinfo {year} {2011})}\BibitemShut {NoStop}%
\bibitem [{\citenamefont {Wouters}\ and\ \citenamefont {van
  Neck}(2014)}]{wout14}%
  \BibitemOpen
  \bibfield  {author} {\bibinfo {author} {\bibfnamefont {S.}~\bibnamefont
  {Wouters}}\ and\ \bibinfo {author} {\bibfnamefont {D.}~\bibnamefont {van
  Neck}},\ }\href@noop {} {\bibfield  {journal} {\bibinfo  {journal} {Eur.
  Phys. J. D}\ }\textbf {\bibinfo {volume} {68}},\ \bibinfo {pages} {272}
  (\bibinfo {year} {2014})}\BibitemShut {NoStop}%
\bibitem [{\citenamefont {Kurashige}(2014)}]{kura14a}%
  \BibitemOpen
  \bibfield  {author} {\bibinfo {author} {\bibfnamefont {Y.}~\bibnamefont
  {Kurashige}},\ }\href@noop {} {\bibfield  {journal} {\bibinfo  {journal}
  {Mol. Phys.}\ }\textbf {\bibinfo {volume} {112}},\ \bibinfo {pages} {1485}
  (\bibinfo {year} {2014})}\BibitemShut {NoStop}%
\bibitem [{\citenamefont {Yanai}\ \emph {et~al.}(2015)\citenamefont {Yanai},
  \citenamefont {Kurashige}, \citenamefont {Mizukami}, \citenamefont
  {Chalupsky}, \citenamefont {Lan},\ and\ \citenamefont {Saitow}}]{yana15}%
  \BibitemOpen
  \bibfield  {author} {\bibinfo {author} {\bibfnamefont {T.}~\bibnamefont
  {Yanai}}, \bibinfo {author} {\bibfnamefont {Y.}~\bibnamefont {Kurashige}},
  \bibinfo {author} {\bibfnamefont {W.}~\bibnamefont {Mizukami}}, \bibinfo
  {author} {\bibfnamefont {J.}~\bibnamefont {Chalupsky}}, \bibinfo {author}
  {\bibfnamefont {T.~N.}\ \bibnamefont {Lan}}, \ and\ \bibinfo {author}
  {\bibfnamefont {M.}~\bibnamefont {Saitow}},\ }\href@noop {} {\bibfield
  {journal} {\bibinfo  {journal} {Int. J. Quantum Chem.}\ }\textbf {\bibinfo
  {volume} {115}},\ \bibinfo {pages} {283} (\bibinfo {year}
  {2015})}\BibitemShut {NoStop}%
\bibitem [{\citenamefont {Szalay}\ \emph {et~al.}(2015)\citenamefont {Szalay},
  \citenamefont {Pfeffer}, \citenamefont {Murg}, \citenamefont {Barcza},
  \citenamefont {Verstraete}, \citenamefont {Schneider},\ and\ \citenamefont
  {Legeza}}]{szal15}%
  \BibitemOpen
  \bibfield  {author} {\bibinfo {author} {\bibfnamefont {S.}~\bibnamefont
  {Szalay}}, \bibinfo {author} {\bibfnamefont {M.}~\bibnamefont {Pfeffer}},
  \bibinfo {author} {\bibfnamefont {V.}~\bibnamefont {Murg}}, \bibinfo {author}
  {\bibfnamefont {G.}~\bibnamefont {Barcza}}, \bibinfo {author} {\bibfnamefont
  {F.}~\bibnamefont {Verstraete}}, \bibinfo {author} {\bibfnamefont
  {R.}~\bibnamefont {Schneider}}, \ and\ \bibinfo {author} {\bibfnamefont
  {{\"O}.}~\bibnamefont {Legeza}},\ }\href@noop {} {\bibfield  {journal}
  {\bibinfo  {journal} {Int. J. Quantum Chem.}\ }\textbf {\bibinfo {volume}
  {115}},\ \bibinfo {pages} {1342} (\bibinfo {year} {2015})}\BibitemShut
  {NoStop}%
\bibitem [{\citenamefont {Knecht}\ \emph {et~al.}(2016)\citenamefont {Knecht},
  \citenamefont {Hedeg{\r{a}}rd}, \citenamefont {Keller}, \citenamefont
  {Kovyrshin}, \citenamefont {Ma}, \citenamefont {Muolo}, \citenamefont
  {Stein},\ and\ \citenamefont {Reiher}}]{knec16}%
  \BibitemOpen
  \bibfield  {author} {\bibinfo {author} {\bibfnamefont {S.}~\bibnamefont
  {Knecht}}, \bibinfo {author} {\bibfnamefont {E.~D.}\ \bibnamefont
  {Hedeg{\r{a}}rd}}, \bibinfo {author} {\bibfnamefont {S.}~\bibnamefont
  {Keller}}, \bibinfo {author} {\bibfnamefont {A.}~\bibnamefont {Kovyrshin}},
  \bibinfo {author} {\bibfnamefont {Y.}~\bibnamefont {Ma}}, \bibinfo {author}
  {\bibfnamefont {A.}~\bibnamefont {Muolo}}, \bibinfo {author} {\bibfnamefont
  {C.~J.}\ \bibnamefont {Stein}}, \ and\ \bibinfo {author} {\bibfnamefont
  {M.}~\bibnamefont {Reiher}},\ }\href@noop {} {\bibfield  {journal} {\bibinfo
  {journal} {Chimia}\ }\textbf {\bibinfo {volume} {70}},\ \bibinfo {pages}
  {244} (\bibinfo {year} {2016})}\BibitemShut {NoStop}%
\bibitem [{\citenamefont {Chan}\ \emph {et~al.}(2016)\citenamefont {Chan},
  \citenamefont {Keselman}, \citenamefont {Nakatani}, \citenamefont {Li},\ and\
  \citenamefont {White}}]{chan16}%
  \BibitemOpen
  \bibfield  {author} {\bibinfo {author} {\bibfnamefont {G.~K.-L.}\
  \bibnamefont {Chan}}, \bibinfo {author} {\bibfnamefont {A.}~\bibnamefont
  {Keselman}}, \bibinfo {author} {\bibfnamefont {N.}~\bibnamefont {Nakatani}},
  \bibinfo {author} {\bibfnamefont {Z.}~\bibnamefont {Li}}, \ and\ \bibinfo
  {author} {\bibfnamefont {S.~R.}\ \bibnamefont {White}},\ }\href@noop {}
  {\bibfield  {journal} {\bibinfo  {journal} {J. Chem. Phys.}\ }\textbf
  {\bibinfo {volume} {145}},\ \bibinfo {pages} {014102} (\bibinfo {year}
  {2016})}\BibitemShut {NoStop}%
\bibitem [{\citenamefont {Zgid}\ and\ \citenamefont {Nooijen}(2008)}]{zgid08}%
  \BibitemOpen
  \bibfield  {author} {\bibinfo {author} {\bibfnamefont {D.}~\bibnamefont
  {Zgid}}\ and\ \bibinfo {author} {\bibfnamefont {M.}~\bibnamefont {Nooijen}},\
  }\href@noop {} {\bibfield  {journal} {\bibinfo  {journal} {J. Chem. Phys.}\
  }\textbf {\bibinfo {volume} {128}},\ \bibinfo {pages} {144116} (\bibinfo
  {year} {2008})}\BibitemShut {NoStop}%
\bibitem [{\citenamefont {Ghosh}\ \emph {et~al.}(2008)\citenamefont {Ghosh},
  \citenamefont {Hachmann}, \citenamefont {Yanai},\ and\ \citenamefont
  {Chan}}]{ghos08}%
  \BibitemOpen
  \bibfield  {author} {\bibinfo {author} {\bibfnamefont {D.}~\bibnamefont
  {Ghosh}}, \bibinfo {author} {\bibfnamefont {J.}~\bibnamefont {Hachmann}},
  \bibinfo {author} {\bibfnamefont {T.}~\bibnamefont {Yanai}}, \ and\ \bibinfo
  {author} {\bibfnamefont {G.~K.-L.}\ \bibnamefont {Chan}},\ }\href@noop {}
  {\bibfield  {journal} {\bibinfo  {journal} {J. Chem. Phys.}\ }\textbf
  {\bibinfo {volume} {128}},\ \bibinfo {pages} {144117} (\bibinfo {year}
  {2008})}\BibitemShut {NoStop}%
\bibitem [{\citenamefont {Knecht}, \citenamefont {Legeza},\ and\ \citenamefont
  {Reiher}(2014)}]{knec14}%
  \BibitemOpen
  \bibfield  {author} {\bibinfo {author} {\bibfnamefont {S.}~\bibnamefont
  {Knecht}}, \bibinfo {author} {\bibfnamefont {O.}~\bibnamefont {Legeza}}, \
  and\ \bibinfo {author} {\bibfnamefont {M.}~\bibnamefont {Reiher}},\
  }\href@noop {} {\bibfield  {journal} {\bibinfo  {journal} {J. Chem. Phys.}\
  }\textbf {\bibinfo {volume} {140}},\ \bibinfo {pages} {041101} (\bibinfo
  {year} {2014})}\BibitemShut {NoStop}%
\bibitem [{\citenamefont {Malmqvist}\ and\ \citenamefont
  {Roos}(1989)}]{malm89}%
  \BibitemOpen
  \bibfield  {author} {\bibinfo {author} {\bibfnamefont {P.-A.}\ \bibnamefont
  {Malmqvist}}\ and\ \bibinfo {author} {\bibfnamefont {B.~O.}\ \bibnamefont
  {Roos}},\ }\href@noop {} {\bibfield  {journal} {\bibinfo  {journal} {Chem.
  Phys. Lett.}\ }\textbf {\bibinfo {volume} {155}},\ \bibinfo {pages} {189}
  (\bibinfo {year} {1989})}\BibitemShut {NoStop}%
\bibitem [{\citenamefont {Malmqvist}(1986)}]{malm86}%
  \BibitemOpen
  \bibfield  {author} {\bibinfo {author} {\bibfnamefont {P.-A.}\ \bibnamefont
  {Malmqvist}},\ }\href@noop {} {\bibfield  {journal} {\bibinfo  {journal}
  {Int. J. Quantum Chem.}\ }\textbf {\bibinfo {volume} {30}},\ \bibinfo {pages}
  {479} (\bibinfo {year} {1986})}\BibitemShut {NoStop}%
\bibitem [{\citenamefont {Olsen}\ \emph {et~al.}(1995)\citenamefont {Olsen},
  \citenamefont {Godefroid}, \citenamefont {J{\"o}nsson}, \citenamefont
  {Malmqvist},\ and\ \citenamefont {Fischer}}]{olse95}%
  \BibitemOpen
  \bibfield  {author} {\bibinfo {author} {\bibfnamefont {J.}~\bibnamefont
  {Olsen}}, \bibinfo {author} {\bibfnamefont {M.~R.}\ \bibnamefont
  {Godefroid}}, \bibinfo {author} {\bibfnamefont {P.}~\bibnamefont
  {J{\"o}nsson}}, \bibinfo {author} {\bibfnamefont {P.-A.}\ \bibnamefont
  {Malmqvist}}, \ and\ \bibinfo {author} {\bibfnamefont {C.~F.}\ \bibnamefont
  {Fischer}},\ }\href@noop {} {\bibfield  {journal} {\bibinfo  {journal} {Phys.
  Rev. E}\ }\textbf {\bibinfo {volume} {52}},\ \bibinfo {pages} {4499}
  (\bibinfo {year} {1995})}\BibitemShut {NoStop}%
\bibitem [{\citenamefont {Olsen}(2015)}]{olse15}%
  \BibitemOpen
  \bibfield  {author} {\bibinfo {author} {\bibfnamefont {J.}~\bibnamefont
  {Olsen}},\ }\href@noop {} {\bibfield  {journal} {\bibinfo  {journal} {J.
  Chem. Phys.}\ }\textbf {\bibinfo {volume} {143}},\ \bibinfo {pages} {114102}
  (\bibinfo {year} {2015})}\BibitemShut {NoStop}%
\bibitem [{\citenamefont {Chen}, \citenamefont {Chen},\ and\ \citenamefont
  {Wu}(2013{\natexlab{a}})}]{chen13b}%
  \BibitemOpen
  \bibfield  {author} {\bibinfo {author} {\bibfnamefont {Z.}~\bibnamefont
  {Chen}}, \bibinfo {author} {\bibfnamefont {X.}~\bibnamefont {Chen}}, \ and\
  \bibinfo {author} {\bibfnamefont {W.}~\bibnamefont {Wu}},\ }\href@noop {}
  {\bibfield  {journal} {\bibinfo  {journal} {J. Chem. Phys.}\ }\textbf
  {\bibinfo {volume} {138}},\ \bibinfo {pages} {164120} (\bibinfo {year}
  {2013}{\natexlab{a}})}\BibitemShut {NoStop}%
\bibitem [{\citenamefont {Olsen}\ \emph {et~al.}(1988)\citenamefont {Olsen},
  \citenamefont {Roos}, \citenamefont {J\o{}rgensen},\ and\ \citenamefont
  {Jensen}}]{olse88}%
  \BibitemOpen
  \bibfield  {author} {\bibinfo {author} {\bibfnamefont {J.}~\bibnamefont
  {Olsen}}, \bibinfo {author} {\bibfnamefont {B.~O.}\ \bibnamefont {Roos}},
  \bibinfo {author} {\bibfnamefont {P.}~\bibnamefont {J\o{}rgensen}}, \ and\
  \bibinfo {author} {\bibfnamefont {H.~J.~A.}\ \bibnamefont {Jensen}},\
  }\href@noop {} {\bibfield  {journal} {\bibinfo  {journal} {J. Chem. Phys.}\
  }\textbf {\bibinfo {volume} {89}},\ \bibinfo {pages} {2185} (\bibinfo {year}
  {1988})}\BibitemShut {NoStop}%
\bibitem [{\citenamefont {Roos}(1992)}]{roos92}%
  \BibitemOpen
  \bibfield  {author} {\bibinfo {author} {\bibfnamefont {B.~O.}\ \bibnamefont
  {Roos}},\ }in\ \href@noop {} {\emph {\bibinfo {booktitle} {Lecture Notes in
  Quantum Chemistry: European Summer School in Quantum Chemistry}}},\ \bibinfo
  {editor} {edited by\ \bibinfo {editor} {\bibfnamefont {B.~O.}\ \bibnamefont
  {Roos}}}\ (\bibinfo  {publisher} {Springer Verlag},\ \bibinfo {address}
  {Berlin Heidelberg},\ \bibinfo {year} {1992})\ pp.\ \bibinfo {pages}
  {177--254}\BibitemShut {NoStop}%
\bibitem [{\citenamefont {Roos}(2008)}]{roos08a}%
  \BibitemOpen
  \bibfield  {author} {\bibinfo {author} {\bibfnamefont {B.~O.}\ \bibnamefont
  {Roos}},\ }in\ \href@noop {} {\emph {\bibinfo {booktitle} {Radiation Induced
  Molecular Phenomena in Nucleic Acids}}},\ \bibinfo {editor} {edited by\
  \bibinfo {editor} {\bibfnamefont {M.~K.}\ \bibnamefont {Shukla}}\ and\
  \bibinfo {editor} {\bibfnamefont {J.}~\bibnamefont {Leszczynski}}}\ (\bibinfo
   {publisher} {Springer},\ \bibinfo {year} {2008})\ pp.\ \bibinfo {pages}
  {125--156}\BibitemShut {NoStop}%
\bibitem [{\citenamefont {Keller}\ \emph {et~al.}(2015)\citenamefont {Keller},
  \citenamefont {Dolfi}, \citenamefont {Troyer},\ and\ \citenamefont
  {Reiher}}]{kell15}%
  \BibitemOpen
  \bibfield  {author} {\bibinfo {author} {\bibfnamefont {S.}~\bibnamefont
  {Keller}}, \bibinfo {author} {\bibfnamefont {M.}~\bibnamefont {Dolfi}},
  \bibinfo {author} {\bibfnamefont {M.}~\bibnamefont {Troyer}}, \ and\ \bibinfo
  {author} {\bibfnamefont {M.}~\bibnamefont {Reiher}},\ }\href@noop {}
  {\bibfield  {journal} {\bibinfo  {journal} {J. Chem. Phys.}\ }\textbf
  {\bibinfo {volume} {143}},\ \bibinfo {pages} {244118} (\bibinfo {year}
  {2015})}\BibitemShut {NoStop}%
\bibitem [{\citenamefont {Keller}\ and\ \citenamefont {Reiher}(2016)}]{kell16}%
  \BibitemOpen
  \bibfield  {author} {\bibinfo {author} {\bibfnamefont {S.}~\bibnamefont
  {Keller}}\ and\ \bibinfo {author} {\bibfnamefont {M.}~\bibnamefont
  {Reiher}},\ }\href@noop {} {\bibfield  {journal} {\bibinfo  {journal} {J.
  Chem. Phys.}\ }\textbf {\bibinfo {volume} {144}},\ \bibinfo {pages} {134101}
  (\bibinfo {year} {2016})}\BibitemShut {NoStop}%
\bibitem [{\citenamefont {Moshinsky}\ and\ \citenamefont
  {Seligman}(1971)}]{mosh71}%
  \BibitemOpen
  \bibfield  {author} {\bibinfo {author} {\bibfnamefont {M.}~\bibnamefont
  {Moshinsky}}\ and\ \bibinfo {author} {\bibfnamefont {T.~H.}\ \bibnamefont
  {Seligman}},\ }\href@noop {} {\bibfield  {journal} {\bibinfo  {journal} {Ann.
  Phys.}\ }\textbf {\bibinfo {volume} {66}},\ \bibinfo {pages} {311} (\bibinfo
  {year} {1971})}\BibitemShut {NoStop}%
\bibitem [{\citenamefont {Helgaker}, \citenamefont {J{\o}rgensen},\ and\
  \citenamefont {Olsen}(2000)}]{helg00}%
  \BibitemOpen
  \bibinfo {editor} {\bibfnamefont {T.}~\bibnamefont {Helgaker}}, \bibinfo
  {editor} {\bibfnamefont {P.}~\bibnamefont {J{\o}rgensen}}, \ and\ \bibinfo
  {editor} {\bibfnamefont {J.}~\bibnamefont {Olsen}},\ eds.,\ \href@noop {}
  {\emph {\bibinfo {title} {{M}olecular {E}lectronic-{S}tructure {T}heory}}},\
  \bibinfo {edition} {1st}\ ed.\ (\bibinfo  {publisher} {Wiley {\&} Sons,
  Chichester (England)},\ \bibinfo {year} {2000})\BibitemShut {NoStop}%
\bibitem [{\citenamefont {L{\"o}wdin}(1950)}]{lowd50}%
  \BibitemOpen
  \bibfield  {author} {\bibinfo {author} {\bibfnamefont {P.-O.}\ \bibnamefont
  {L{\"o}wdin}},\ }\href@noop {} {\bibfield  {journal} {\bibinfo  {journal} {J.
  Chem. Phys.}\ }\textbf {\bibinfo {volume} {18}},\ \bibinfo {pages} {365}
  (\bibinfo {year} {1950})}\BibitemShut {NoStop}%
\bibitem [{\citenamefont {Heitler}\ and\ \citenamefont
  {London}(1927)}]{heit27}%
  \BibitemOpen
  \bibfield  {author} {\bibinfo {author} {\bibfnamefont {W.}~\bibnamefont
  {Heitler}}\ and\ \bibinfo {author} {\bibfnamefont {F.}~\bibnamefont
  {London}},\ }\href@noop {} {\bibfield  {journal} {\bibinfo  {journal} {Z.
  Phys.}\ }\textbf {\bibinfo {volume} {44}},\ \bibinfo {pages} {455} (\bibinfo
  {year} {1927})}\BibitemShut {NoStop}%
\bibitem [{\citenamefont {L{\"o}wdin}(1991)}]{lowd91}%
  \BibitemOpen
  \bibfield  {author} {\bibinfo {author} {\bibfnamefont {P.}~\bibnamefont
  {L{\"o}wdin}},\ }\href@noop {} {\bibfield  {journal} {\bibinfo  {journal} {J.
  Mol. Struct. (Theochem)}\ }\textbf {\bibinfo {volume} {229}},\ \bibinfo
  {pages} {1} (\bibinfo {year} {1991})}\BibitemShut {NoStop}%
\bibitem [{\citenamefont {Wu}\ \emph {et~al.}(2011)\citenamefont {Wu},
  \citenamefont {Su}, \citenamefont {Shaik},\ and\ \citenamefont
  {Hiberty}}]{vb_review2011}%
  \BibitemOpen
  \bibfield  {author} {\bibinfo {author} {\bibfnamefont {W.}~\bibnamefont
  {Wu}}, \bibinfo {author} {\bibfnamefont {P.}~\bibnamefont {Su}}, \bibinfo
  {author} {\bibfnamefont {S.}~\bibnamefont {Shaik}}, \ and\ \bibinfo {author}
  {\bibfnamefont {P.~C.}\ \bibnamefont {Hiberty}},\ }\href@noop {} {\bibfield
  {journal} {\bibinfo  {journal} {Chem. Rev.}\ }\textbf {\bibinfo {volume}
  {111}},\ \bibinfo {pages} {7557–7593} (\bibinfo {year} {2011})}\BibitemShut
  {NoStop}%
\bibitem [{\citenamefont {Rashid}\ and\ \citenamefont {van
  Lenthe}(2013)}]{vanl13}%
  \BibitemOpen
  \bibfield  {author} {\bibinfo {author} {\bibfnamefont {Z.}~\bibnamefont
  {Rashid}}\ and\ \bibinfo {author} {\bibfnamefont {J.~H.}\ \bibnamefont {van
  Lenthe}},\ }\href@noop {} {\bibfield  {journal} {\bibinfo  {journal} {J.
  Chem. Phys.}\ }\textbf {\bibinfo {volume} {138}},\ \bibinfo {pages} {054105}
  (\bibinfo {year} {2013})}\BibitemShut {NoStop}%
\bibitem [{\citenamefont {Chen}, \citenamefont {Chen},\ and\ \citenamefont
  {Wu}(2013{\natexlab{b}})}]{chen13a}%
  \BibitemOpen
  \bibfield  {author} {\bibinfo {author} {\bibfnamefont {Z.}~\bibnamefont
  {Chen}}, \bibinfo {author} {\bibfnamefont {X.}~\bibnamefont {Chen}}, \ and\
  \bibinfo {author} {\bibfnamefont {W.}~\bibnamefont {Wu}},\ }\href@noop {}
  {\bibfield  {journal} {\bibinfo  {journal} {J. Chem. Phys.}\ }\textbf
  {\bibinfo {volume} {138}},\ \bibinfo {pages} {164119} (\bibinfo {year}
  {2013}{\natexlab{b}})}\BibitemShut {NoStop}%
\bibitem [{\citenamefont {Thom}\ and\ \citenamefont
  {Head-Gordon}(2009)}]{thom09}%
  \BibitemOpen
  \bibfield  {author} {\bibinfo {author} {\bibfnamefont {A.~J.~W.}\
  \bibnamefont {Thom}}\ and\ \bibinfo {author} {\bibfnamefont {M.}~\bibnamefont
  {Head-Gordon}},\ }\href@noop {} {\bibfield  {journal} {\bibinfo  {journal}
  {J. Chem. Phys.}\ }\textbf {\bibinfo {volume} {131}},\ \bibinfo {pages}
  {124113} (\bibinfo {year} {2009})}\BibitemShut {NoStop}%
\bibitem [{\citenamefont {Sundstrom}\ and\ \citenamefont
  {Head-Gordon}(2014)}]{sund14}%
  \BibitemOpen
  \bibfield  {author} {\bibinfo {author} {\bibfnamefont {E.~J.}\ \bibnamefont
  {Sundstrom}}\ and\ \bibinfo {author} {\bibfnamefont {M.}~\bibnamefont
  {Head-Gordon}},\ }\href@noop {} {\bibfield  {journal} {\bibinfo  {journal}
  {J. Chem. Phys.}\ }\textbf {\bibinfo {volume} {140}},\ \bibinfo {pages}
  {114103} (\bibinfo {year} {2014})}\BibitemShut {NoStop}%
\bibitem [{\citenamefont {L\"{o}wdin}(1955)}]{lowd55}%
  \BibitemOpen
  \bibfield  {author} {\bibinfo {author} {\bibfnamefont {P.-O.}\ \bibnamefont
  {L\"{o}wdin}},\ }\href@noop {} {\bibfield  {journal} {\bibinfo  {journal}
  {Phys. Rev.}\ }\textbf {\bibinfo {volume} {97}},\ \bibinfo {pages} {1474}
  (\bibinfo {year} {1955})}\BibitemShut {NoStop}%
\bibitem [{\citenamefont {Prosser}\ and\ \citenamefont
  {Hagstrom}(1968)}]{pros68}%
  \BibitemOpen
  \bibfield  {author} {\bibinfo {author} {\bibfnamefont {F.}~\bibnamefont
  {Prosser}}\ and\ \bibinfo {author} {\bibfnamefont {S.}~\bibnamefont
  {Hagstrom}},\ }\href@noop {} {\bibfield  {journal} {\bibinfo  {journal} {Int.
  J. Quantum Chem.}\ }\textbf {\bibinfo {volume} {2}},\ \bibinfo {pages} {89}
  (\bibinfo {year} {1968})}\BibitemShut {NoStop}%
\bibitem [{\citenamefont {Weltin}(1976)}]{welt76}%
  \BibitemOpen
  \bibfield  {author} {\bibinfo {author} {\bibfnamefont {E.~E.}\ \bibnamefont
  {Weltin}},\ }\href@noop {} {\bibfield  {journal} {\bibinfo  {journal} {Int.
  J. Quantum Chem.}\ }\textbf {\bibinfo {volume} {10}},\ \bibinfo {pages} {163}
  (\bibinfo {year} {1976})}\BibitemShut {NoStop}%
\bibitem [{\citenamefont {Dalgaard}(1983)}]{dalg83}%
  \BibitemOpen
  \bibfield  {author} {\bibinfo {author} {\bibfnamefont {E.}~\bibnamefont
  {Dalgaard}},\ }\href@noop {} {\bibfield  {journal} {\bibinfo  {journal}
  {Theoret. Chim. Acta}\ }\textbf {\bibinfo {volume} {64}},\ \bibinfo {pages}
  {181} (\bibinfo {year} {1983})}\BibitemShut {NoStop}%
\bibitem [{\citenamefont {Verbeek}\ and\ \citenamefont {van
  Lenthe}(1991{\natexlab{a}})}]{verb91}%
  \BibitemOpen
  \bibfield  {author} {\bibinfo {author} {\bibfnamefont {J.}~\bibnamefont
  {Verbeek}}\ and\ \bibinfo {author} {\bibfnamefont {J.~H.}\ \bibnamefont {van
  Lenthe}},\ }\href@noop {} {\bibfield  {journal} {\bibinfo  {journal} {Int. J.
  Quantum Chem.}\ }\textbf {\bibinfo {volume} {40}},\ \bibinfo {pages} {201}
  (\bibinfo {year} {1991}{\natexlab{a}})}\BibitemShut {NoStop}%
\bibitem [{\citenamefont {Verbeek}\ and\ \citenamefont {van
  Lenthe}(1991{\natexlab{b}})}]{vanl91}%
  \BibitemOpen
  \bibfield  {author} {\bibinfo {author} {\bibfnamefont {J.}~\bibnamefont
  {Verbeek}}\ and\ \bibinfo {author} {\bibfnamefont {J.~H.}\ \bibnamefont {van
  Lenthe}},\ }\href@noop {} {\bibfield  {journal} {\bibinfo  {journal} {J. Mol.
  Struct: Theochem}\ }\textbf {\bibinfo {volume} {229}},\ \bibinfo {pages}
  {115} (\bibinfo {year} {1991}{\natexlab{b}})}\BibitemShut {NoStop}%
\bibitem [{\citenamefont {Koch}\ and\ \citenamefont {Dalgaard}(1993)}]{koch93}%
  \BibitemOpen
  \bibfield  {author} {\bibinfo {author} {\bibfnamefont {H.}~\bibnamefont
  {Koch}}\ and\ \bibinfo {author} {\bibfnamefont {E.}~\bibnamefont
  {Dalgaard}},\ }\href@noop {} {\bibfield  {journal} {\bibinfo  {journal}
  {Chem. Phys. Lett.}\ }\textbf {\bibinfo {volume} {212}},\ \bibinfo {pages}
  {193} (\bibinfo {year} {1993})}\BibitemShut {NoStop}%
\bibitem [{\citenamefont {Amovilli}(1997)}]{amov97}%
  \BibitemOpen
  \bibfield  {author} {\bibinfo {author} {\bibfnamefont {C.}~\bibnamefont
  {Amovilli}},\ }in\ \href@noop {} {\emph {\bibinfo {booktitle} {Quantum
  Systems in Quantum Chemistry and Physics}}},\ \bibinfo {editor} {edited by\
  \bibinfo {editor} {\bibfnamefont {R.}~\bibnamefont {McWeeny}}, \bibinfo
  {editor} {\bibfnamefont {J.}~\bibnamefont {Maruani}}, \bibinfo {editor}
  {\bibfnamefont {Y.~G.}\ \bibnamefont {Smeyers}}, \ and\ \bibinfo {editor}
  {\bibfnamefont {S.}~\bibnamefont {Wilson}}}\ (\bibinfo  {publisher} {Kluwer
  Academics Publishers},\ \bibinfo {year} {1997})\ pp.\ \bibinfo {pages}
  {343--347}\BibitemShut {NoStop}%
\bibitem [{\citenamefont {Dijkstra}\ and\ \citenamefont {van
  Lenthe}(1998)}]{vanl98}%
  \BibitemOpen
  \bibfield  {author} {\bibinfo {author} {\bibfnamefont {F.}~\bibnamefont
  {Dijkstra}}\ and\ \bibinfo {author} {\bibfnamefont {J.~H.}\ \bibnamefont {van
  Lenthe}},\ }\href@noop {} {\bibfield  {journal} {\bibinfo  {journal} {Int. J.
  Quantum Chem.}\ }\textbf {\bibinfo {volume} {67}},\ \bibinfo {pages} {77}
  (\bibinfo {year} {1998})}\BibitemShut {NoStop}%
\bibitem [{\citenamefont {Slater}(1929)}]{slat29}%
  \BibitemOpen
  \bibfield  {author} {\bibinfo {author} {\bibfnamefont {J.}~\bibnamefont
  {Slater}},\ }\href@noop {} {\bibfield  {journal} {\bibinfo  {journal} {Phys.
  Rev.}\ }\textbf {\bibinfo {volume} {34}},\ \bibinfo {pages} {1293} (\bibinfo
  {year} {1929})}\BibitemShut {NoStop}%
\bibitem [{\citenamefont {Condon}(1930)}]{cond30}%
  \BibitemOpen
  \bibfield  {author} {\bibinfo {author} {\bibfnamefont {E.}~\bibnamefont
  {Condon}},\ }\href@noop {} {\bibfield  {journal} {\bibinfo  {journal} {Phys.
  Rev.}\ }\textbf {\bibinfo {volume} {36}},\ \bibinfo {pages} {1121} (\bibinfo
  {year} {1930})}\BibitemShut {NoStop}%
\bibitem [{\citenamefont {McDouall}(1992)}]{doua92}%
  \BibitemOpen
  \bibfield  {author} {\bibinfo {author} {\bibfnamefont {J.~J.~W.}\
  \bibnamefont {McDouall}},\ }\href@noop {} {\bibfield  {journal} {\bibinfo
  {journal} {Theor. Chim. Acta}\ }\textbf {\bibinfo {volume} {83}},\ \bibinfo
  {pages} {339} (\bibinfo {year} {1992})}\BibitemShut {NoStop}%
\bibitem [{\citenamefont {Malmqvist}, \citenamefont {Roos},\ and\ \citenamefont
  {Schimmelpfennig}(2002)}]{malm02}%
  \BibitemOpen
  \bibfield  {author} {\bibinfo {author} {\bibfnamefont {P.-A.}\ \bibnamefont
  {Malmqvist}}, \bibinfo {author} {\bibfnamefont {B.~O.}\ \bibnamefont {Roos}},
  \ and\ \bibinfo {author} {\bibfnamefont {B.}~\bibnamefont
  {Schimmelpfennig}},\ }\href@noop {} {\bibfield  {journal} {\bibinfo
  {journal} {Chem. Phys. Lett.}\ }\textbf {\bibinfo {volume} {357}},\ \bibinfo
  {pages} {230} (\bibinfo {year} {2002})}\BibitemShut {NoStop}%
\bibitem [{\citenamefont {Jordan}\ and\ \citenamefont {Wigner}(1928)}]{jord28}%
  \BibitemOpen
  \bibfield  {author} {\bibinfo {author} {\bibfnamefont {P.}~\bibnamefont
  {Jordan}}\ and\ \bibinfo {author} {\bibfnamefont {E.}~\bibnamefont
  {Wigner}},\ }\href@noop {} {\bibfield  {journal} {\bibinfo  {journal} {Z.
  Phys.}\ }\textbf {\bibinfo {volume} {47}},\ \bibinfo {pages} {631} (\bibinfo
  {year} {1928})}\BibitemShut {NoStop}%
\bibitem [{\citenamefont {Aquilante}\ \emph
  {et~al.}(2015{\natexlab{b}})\citenamefont {Aquilante}, \citenamefont
  {Autschbach}, \citenamefont {Carlson}, \citenamefont {Chibotaru},
  \citenamefont {Delcey}, \citenamefont {De~Vico}, \citenamefont
  {Fdez.~Galván}, \citenamefont {Ferr{\'e}}, \citenamefont {Frutos},
  \citenamefont {Gagliardi}, \citenamefont {Garavelli}, \citenamefont
  {Giussani}, \citenamefont {Hoyer}, \citenamefont {Li~Manni}, \citenamefont
  {Lischka}, \citenamefont {Ma}, \citenamefont {Malmqvist}, \citenamefont
  {M{\"u}ller}, \citenamefont {Nenov}, \citenamefont {Olivucci}, \citenamefont
  {Pedersen}, \citenamefont {Peng}, \citenamefont {Plasser}, \citenamefont
  {Pritchard}, \citenamefont {Reiher}, \citenamefont {Rivalta}, \citenamefont
  {Schapiro}, \citenamefont {Segarra-Mart{\'i}}, \citenamefont {Stenrup},
  \citenamefont {Truhlar}, \citenamefont {Ungur}, \citenamefont {Valentini},
  \citenamefont {Vancoillie}, \citenamefont {Veryazov}, \citenamefont
  {Vysotskiy}, \citenamefont {Weingart}, \citenamefont {Zapata},\ and\
  \citenamefont {Lindh}}]{molc15}%
  \BibitemOpen
  \bibfield  {author} {\bibinfo {author} {\bibfnamefont {F.}~\bibnamefont
  {Aquilante}}, \bibinfo {author} {\bibfnamefont {J.}~\bibnamefont
  {Autschbach}}, \bibinfo {author} {\bibfnamefont {R.~K.}\ \bibnamefont
  {Carlson}}, \bibinfo {author} {\bibfnamefont {L.~F.}\ \bibnamefont
  {Chibotaru}}, \bibinfo {author} {\bibfnamefont {M.~G.}\ \bibnamefont
  {Delcey}}, \bibinfo {author} {\bibfnamefont {L.}~\bibnamefont {De~Vico}},
  \bibinfo {author} {\bibfnamefont {I.}~\bibnamefont {Fdez.~Galván}}, \bibinfo
  {author} {\bibfnamefont {N.}~\bibnamefont {Ferr{\'e}}}, \bibinfo {author}
  {\bibfnamefont {L.~M.}\ \bibnamefont {Frutos}}, \bibinfo {author}
  {\bibfnamefont {L.}~\bibnamefont {Gagliardi}}, \bibinfo {author}
  {\bibfnamefont {M.}~\bibnamefont {Garavelli}}, \bibinfo {author}
  {\bibfnamefont {A.}~\bibnamefont {Giussani}}, \bibinfo {author}
  {\bibfnamefont {C.~E.}\ \bibnamefont {Hoyer}}, \bibinfo {author}
  {\bibfnamefont {G.}~\bibnamefont {Li~Manni}}, \bibinfo {author}
  {\bibfnamefont {H.}~\bibnamefont {Lischka}}, \bibinfo {author} {\bibfnamefont
  {D.}~\bibnamefont {Ma}}, \bibinfo {author} {\bibfnamefont {P.~{\AA}.}\
  \bibnamefont {Malmqvist}}, \bibinfo {author} {\bibfnamefont {T.}~\bibnamefont
  {M{\"u}ller}}, \bibinfo {author} {\bibfnamefont {A.}~\bibnamefont {Nenov}},
  \bibinfo {author} {\bibfnamefont {M.}~\bibnamefont {Olivucci}}, \bibinfo
  {author} {\bibfnamefont {T.~B.}\ \bibnamefont {Pedersen}}, \bibinfo {author}
  {\bibfnamefont {D.}~\bibnamefont {Peng}}, \bibinfo {author} {\bibfnamefont
  {F.}~\bibnamefont {Plasser}}, \bibinfo {author} {\bibfnamefont
  {B.}~\bibnamefont {Pritchard}}, \bibinfo {author} {\bibfnamefont
  {M.}~\bibnamefont {Reiher}}, \bibinfo {author} {\bibfnamefont
  {I.}~\bibnamefont {Rivalta}}, \bibinfo {author} {\bibfnamefont
  {I.}~\bibnamefont {Schapiro}}, \bibinfo {author} {\bibfnamefont
  {J.}~\bibnamefont {Segarra-Mart{\'i}}}, \bibinfo {author} {\bibfnamefont
  {M.}~\bibnamefont {Stenrup}}, \bibinfo {author} {\bibfnamefont {D.~G.}\
  \bibnamefont {Truhlar}}, \bibinfo {author} {\bibfnamefont {L.}~\bibnamefont
  {Ungur}}, \bibinfo {author} {\bibfnamefont {A.}~\bibnamefont {Valentini}},
  \bibinfo {author} {\bibfnamefont {S.}~\bibnamefont {Vancoillie}}, \bibinfo
  {author} {\bibfnamefont {V.}~\bibnamefont {Veryazov}}, \bibinfo {author}
  {\bibfnamefont {V.~P.}\ \bibnamefont {Vysotskiy}}, \bibinfo {author}
  {\bibfnamefont {O.}~\bibnamefont {Weingart}}, \bibinfo {author}
  {\bibfnamefont {F.}~\bibnamefont {Zapata}}, \ and\ \bibinfo {author}
  {\bibfnamefont {R.}~\bibnamefont {Lindh}},\ }\href@noop {} {\bibfield
  {journal} {\bibinfo  {journal} {J. Comp. Chem.}\ }\textbf {\bibinfo {volume}
  {37}},\ \bibinfo {pages} {506} (\bibinfo {year}
  {2015}{\natexlab{b}})}\BibitemShut {NoStop}%
\bibitem [{\citenamefont {Douglas}\ and\ \citenamefont {Kroll}(1974)}]{doug74}%
  \BibitemOpen
  \bibfield  {author} {\bibinfo {author} {\bibfnamefont {M.}~\bibnamefont
  {Douglas}}\ and\ \bibinfo {author} {\bibfnamefont {N.~M.}\ \bibnamefont
  {Kroll}},\ }\href@noop {} {\bibfield  {journal} {\bibinfo  {journal} {Ann.
  Phys.}\ }\textbf {\bibinfo {volume} {82}},\ \bibinfo {pages} {89} (\bibinfo
  {year} {1974})}\BibitemShut {NoStop}%
\bibitem [{\citenamefont {Hess}(1986)}]{hess86}%
  \BibitemOpen
  \bibfield  {author} {\bibinfo {author} {\bibfnamefont {B.~A.}\ \bibnamefont
  {Hess}},\ }\href@noop {} {\bibfield  {journal} {\bibinfo  {journal} {Phys.
  Rev. A}\ }\textbf {\bibinfo {volume} {33}},\ \bibinfo {pages} {3742}
  (\bibinfo {year} {1986})}\BibitemShut {NoStop}%
\bibitem [{\citenamefont {Reiher}(2006)}]{reih06}%
  \BibitemOpen
  \bibfield  {author} {\bibinfo {author} {\bibfnamefont {M.}~\bibnamefont
  {Reiher}},\ }\href@noop {} {\bibfield  {journal} {\bibinfo  {journal} {Theor.
  Chem. Acc.}\ }\textbf {\bibinfo {volume} {116}},\ \bibinfo {pages} {241}
  (\bibinfo {year} {2006})}\BibitemShut {NoStop}%
\bibitem [{jen()}]{jens05}%
  \BibitemOpen
  \href@noop {} {}\bibinfo {note} {{H}. J. {Aa}. Jensen, oral presentation
  given at the {REHE} conference in M{\"u}hlheim (Germany), 2005.}\BibitemShut
  {Stop}%
\bibitem [{\citenamefont {Kutzelnigg}\ and\ \citenamefont
  {Liu}(2005)}]{kutz05}%
  \BibitemOpen
  \bibfield  {author} {\bibinfo {author} {\bibfnamefont {W.}~\bibnamefont
  {Kutzelnigg}}\ and\ \bibinfo {author} {\bibfnamefont {W.}~\bibnamefont
  {Liu}},\ }\href@noop {} {\bibfield  {journal} {\bibinfo  {journal} {J. Chem.
  Phys.}\ }\textbf {\bibinfo {volume} {123}},\ \bibinfo {pages} {241102}
  (\bibinfo {year} {2005})}\BibitemShut {NoStop}%
\bibitem [{\citenamefont {Saue}(2011)}]{saue11}%
  \BibitemOpen
  \bibfield  {author} {\bibinfo {author} {\bibfnamefont {T.}~\bibnamefont
  {Saue}},\ }\href@noop {} {\bibfield  {journal} {\bibinfo  {journal}
  {ChemPhysChem}\ }\textbf {\bibinfo {volume} {12}},\ \bibinfo {pages} {3077}
  (\bibinfo {year} {2011})}\BibitemShut {NoStop}%
\bibitem [{\citenamefont {Peng}\ and\ \citenamefont {Reiher}(2012)}]{peng12}%
  \BibitemOpen
  \bibfield  {author} {\bibinfo {author} {\bibfnamefont {D.}~\bibnamefont
  {Peng}}\ and\ \bibinfo {author} {\bibfnamefont {M.}~\bibnamefont {Reiher}},\
  }\href@noop {} {\bibfield  {journal} {\bibinfo  {journal} {Theor. Chem.
  Acc.}\ }\textbf {\bibinfo {volume} {131}},\ \bibinfo {pages} {3} (\bibinfo
  {year} {2012})}\BibitemShut {NoStop}%
\bibitem [{\citenamefont {He{\ss}}\ \emph {et~al.}(1996)\citenamefont
  {He{\ss}}, \citenamefont {Marian}, \citenamefont {Wahlgren},\ and\
  \citenamefont {Gropen}}]{hess96}%
  \BibitemOpen
  \bibfield  {author} {\bibinfo {author} {\bibfnamefont {B.~A.}\ \bibnamefont
  {He{\ss}}}, \bibinfo {author} {\bibfnamefont {C.~M.}\ \bibnamefont {Marian}},
  \bibinfo {author} {\bibfnamefont {U.}~\bibnamefont {Wahlgren}}, \ and\
  \bibinfo {author} {\bibfnamefont {O.}~\bibnamefont {Gropen}},\ }\href@noop {}
  {\bibfield  {journal} {\bibinfo  {journal} {Chem. Phys. Lett.}\ }\textbf
  {\bibinfo {volume} {251}},\ \bibinfo {pages} {365} (\bibinfo {year}
  {1996})}\BibitemShut {NoStop}%
\bibitem [{\citenamefont {Tatchen}\ and\ \citenamefont
  {Marian}(1999)}]{tatc99}%
  \BibitemOpen
  \bibfield  {author} {\bibinfo {author} {\bibfnamefont {J.}~\bibnamefont
  {Tatchen}}\ and\ \bibinfo {author} {\bibfnamefont {C.~M.}\ \bibnamefont
  {Marian}},\ }\href@noop {} {\bibfield  {journal} {\bibinfo  {journal} {Chem.
  Phys. Lett.}\ }\textbf {\bibinfo {volume} {313}},\ \bibinfo {pages} {351}
  (\bibinfo {year} {1999})}\BibitemShut {NoStop}%
\bibitem [{amf()}]{amfi96}%
  \BibitemOpen
  \href@noop {} {}\bibinfo {note} {B. Schimmelpfennig, {AMFI is an atomic
  mean-field spin-orbit integral program, University of Stockholm,
  1996.}}\BibitemShut {Stop}%
\bibitem [{\citenamefont {Rose}(1995)}]{rose95}%
  \BibitemOpen
  \bibfield  {author} {\bibinfo {author} {\bibfnamefont {M.~E.}\ \bibnamefont
  {Rose}},\ }\href@noop {} {\emph {\bibinfo {title} {Elementary Theory of
  Angular Momentum}}}\ (\bibinfo  {publisher} {Dover Publications},\ \bibinfo
  {year} {1995})\BibitemShut {NoStop}%
\bibitem [{\citenamefont {Kurashige}\ and\ \citenamefont
  {Yanai}(2011)}]{kura11}%
  \BibitemOpen
  \bibfield  {author} {\bibinfo {author} {\bibfnamefont {Y.}~\bibnamefont
  {Kurashige}}\ and\ \bibinfo {author} {\bibfnamefont {T.}~\bibnamefont
  {Yanai}},\ }\href@noop {} {\bibfield  {journal} {\bibinfo  {journal} {J.
  Chem. Phys.}\ }\textbf {\bibinfo {volume} {135}},\ \bibinfo {pages} {094104}
  (\bibinfo {year} {2011})}\BibitemShut {NoStop}%
\bibitem [{\citenamefont {Kurashige}\ \emph {et~al.}(2014)\citenamefont
  {Kurashige}, \citenamefont {Chalupsky}, \citenamefont {Lan},\ and\
  \citenamefont {Yanai}}]{kura14}%
  \BibitemOpen
  \bibfield  {author} {\bibinfo {author} {\bibfnamefont {Y.}~\bibnamefont
  {Kurashige}}, \bibinfo {author} {\bibfnamefont {J.}~\bibnamefont
  {Chalupsky}}, \bibinfo {author} {\bibfnamefont {T.~N.}\ \bibnamefont {Lan}},
  \ and\ \bibinfo {author} {\bibfnamefont {T.}~\bibnamefont {Yanai}},\
  }\href@noop {} {\bibfield  {journal} {\bibinfo  {journal} {J. Chem. Phys.}\
  }\textbf {\bibinfo {volume} {141}},\ \bibinfo {pages} {174111} (\bibinfo
  {year} {2014})}\BibitemShut {NoStop}%
\bibitem [{\citenamefont {Wouters}, \citenamefont {Van~Speybroeck},\ and\
  \citenamefont {Van~Neck}(2016)}]{wout16b}%
  \BibitemOpen
  \bibfield  {author} {\bibinfo {author} {\bibfnamefont {S.}~\bibnamefont
  {Wouters}}, \bibinfo {author} {\bibfnamefont {V.}~\bibnamefont
  {Van~Speybroeck}}, \ and\ \bibinfo {author} {\bibfnamefont {D.}~\bibnamefont
  {Van~Neck}},\ }\href@noop {} {\bibfield  {journal} {\bibinfo  {journal} {J.
  Chem. Phys.}\ }\textbf {\bibinfo {volume} {145}},\ \bibinfo {pages} {054120}
  (\bibinfo {year} {2016})}\BibitemShut {NoStop}%
\bibitem [{\citenamefont {Sharma}\ and\ \citenamefont {Chan}(2014)}]{shar14}%
  \BibitemOpen
  \bibfield  {author} {\bibinfo {author} {\bibfnamefont {S.}~\bibnamefont
  {Sharma}}\ and\ \bibinfo {author} {\bibfnamefont {G.~K.-L.}\ \bibnamefont
  {Chan}},\ }\href@noop {} {\bibfield  {journal} {\bibinfo  {journal} {J. Chem.
  Phys.}\ }\textbf {\bibinfo {volume} {141}},\ \bibinfo {pages} {111101}
  (\bibinfo {year} {2014})}\BibitemShut {NoStop}%
\bibitem [{\citenamefont {Sokolov}\ and\ \citenamefont {Chan}(2016)}]{soko16}%
  \BibitemOpen
  \bibfield  {author} {\bibinfo {author} {\bibfnamefont {A.~Y.}\ \bibnamefont
  {Sokolov}}\ and\ \bibinfo {author} {\bibfnamefont {G.~K.-L.}\ \bibnamefont
  {Chan}},\ }\href@noop {} {\bibfield  {journal} {\bibinfo  {journal} {J. Chem.
  Phys.}\ }\textbf {\bibinfo {volume} {144}},\ \bibinfo {pages} {064102}
  (\bibinfo {year} {2016})}\BibitemShut {NoStop}%
\bibitem [{\citenamefont {Roemelt}, \citenamefont {Guo},\ and\ \citenamefont
  {Chan}(2016)}]{roem16}%
  \BibitemOpen
  \bibfield  {author} {\bibinfo {author} {\bibfnamefont {M.}~\bibnamefont
  {Roemelt}}, \bibinfo {author} {\bibfnamefont {S.}~\bibnamefont {Guo}}, \ and\
  \bibinfo {author} {\bibfnamefont {G.~K.-L.}\ \bibnamefont {Chan}},\
  }\href@noop {} {\bibfield  {journal} {\bibinfo  {journal} {J. Chem. Phys.}\
  }\textbf {\bibinfo {volume} {144}},\ \bibinfo {pages} {204113} (\bibinfo
  {year} {2016})}\BibitemShut {NoStop}%
\bibitem [{\citenamefont {Guo}\ \emph {et~al.}(2016)\citenamefont {Guo},
  \citenamefont {Watson}, \citenamefont {Hu}, \citenamefont {Sun},\ and\
  \citenamefont {Chan}}]{guos16}%
  \BibitemOpen
  \bibfield  {author} {\bibinfo {author} {\bibfnamefont {S.}~\bibnamefont
  {Guo}}, \bibinfo {author} {\bibfnamefont {M.~A.}\ \bibnamefont {Watson}},
  \bibinfo {author} {\bibfnamefont {W.}~\bibnamefont {Hu}}, \bibinfo {author}
  {\bibfnamefont {Q.}~\bibnamefont {Sun}}, \ and\ \bibinfo {author}
  {\bibfnamefont {G.~K.-L.}\ \bibnamefont {Chan}},\ }\href@noop {} {\bibfield
  {journal} {\bibinfo  {journal} {J. Chem. Theory Comput.}\ }\textbf {\bibinfo
  {volume} {12}},\ \bibinfo {pages} {1583} (\bibinfo {year}
  {2016})}\BibitemShut {NoStop}%
\bibitem [{\citenamefont {Freitag}\ \emph {et~al.}(2016)\citenamefont
  {Freitag}, \citenamefont {Knecht}, \citenamefont {Angeli},\ and\
  \citenamefont {Reiher}}]{frei16}%
  \BibitemOpen
  \bibfield  {author} {\bibinfo {author} {\bibfnamefont {L.}~\bibnamefont
  {Freitag}}, \bibinfo {author} {\bibfnamefont {S.}~\bibnamefont {Knecht}},
  \bibinfo {author} {\bibfnamefont {C.}~\bibnamefont {Angeli}}, \ and\ \bibinfo
  {author} {\bibfnamefont {M.}~\bibnamefont {Reiher}},\ }\href@noop {}
  {\bibfield  {journal} {\bibinfo  {journal} {J. Chem. Theory Comput.}\ }
  (\bibinfo {year} {2016})},\ \bibinfo {note} {submitted,
  arXiv:1608.02006}\BibitemShut {NoStop}%
\bibitem [{\citenamefont {Rota}\ \emph {et~al.}(2011)\citenamefont {Rota},
  \citenamefont {Knecht}, \citenamefont {Fleig}, \citenamefont {Ganyushin},
  \citenamefont {Saue}, \citenamefont {Neese},\ and\ \citenamefont
  {Bolvin}}]{rota11}%
  \BibitemOpen
  \bibfield  {author} {\bibinfo {author} {\bibfnamefont {J.-B.}\ \bibnamefont
  {Rota}}, \bibinfo {author} {\bibfnamefont {S.}~\bibnamefont {Knecht}},
  \bibinfo {author} {\bibfnamefont {T.}~\bibnamefont {Fleig}}, \bibinfo
  {author} {\bibfnamefont {D.}~\bibnamefont {Ganyushin}}, \bibinfo {author}
  {\bibfnamefont {T.}~\bibnamefont {Saue}}, \bibinfo {author} {\bibfnamefont
  {F.}~\bibnamefont {Neese}}, \ and\ \bibinfo {author} {\bibfnamefont
  {H.}~\bibnamefont {Bolvin}},\ }\href@noop {} {\bibfield  {journal} {\bibinfo
  {journal} {J. Chem. Phys.}\ }\textbf {\bibinfo {volume} {135}},\ \bibinfo
  {pages} {114106} (\bibinfo {year} {2011})}\BibitemShut {NoStop}%
\bibitem [{\citenamefont {Roos}\ \emph {et~al.}(2004)\citenamefont {Roos},
  \citenamefont {Lindh}, \citenamefont {Malmqvist}, \citenamefont {Veryazov},\
  and\ \citenamefont {Widmark}}]{anoM04}%
  \BibitemOpen
  \bibfield  {author} {\bibinfo {author} {\bibfnamefont {B.~O.}\ \bibnamefont
  {Roos}}, \bibinfo {author} {\bibfnamefont {R.}~\bibnamefont {Lindh}},
  \bibinfo {author} {\bibfnamefont {P.-{\AA}.}\ \bibnamefont {Malmqvist}},
  \bibinfo {author} {\bibfnamefont {V.}~\bibnamefont {Veryazov}}, \ and\
  \bibinfo {author} {\bibfnamefont {P.-O.}\ \bibnamefont {Widmark}},\
  }\href@noop {} {\bibfield  {journal} {\bibinfo  {journal} {J. Phys. Chem. A}\
  }\textbf {\bibinfo {volume} {108}},\ \bibinfo {pages} {2851} (\bibinfo {year}
  {2004})}\BibitemShut {NoStop}%
\bibitem [{\citenamefont {Roos}\ \emph {et~al.}(2005)\citenamefont {Roos},
  \citenamefont {Lindh}, \citenamefont {Malmqvist}, \citenamefont {Veryazov},\
  and\ \citenamefont {Widmark}}]{anoA05}%
  \BibitemOpen
  \bibfield  {author} {\bibinfo {author} {\bibfnamefont {B.~O.}\ \bibnamefont
  {Roos}}, \bibinfo {author} {\bibfnamefont {R.}~\bibnamefont {Lindh}},
  \bibinfo {author} {\bibfnamefont {P.-{\AA}.}\ \bibnamefont {Malmqvist}},
  \bibinfo {author} {\bibfnamefont {V.}~\bibnamefont {Veryazov}}, \ and\
  \bibinfo {author} {\bibfnamefont {P.-O.}\ \bibnamefont {Widmark}},\
  }\href@noop {} {\bibfield  {journal} {\bibinfo  {journal} {Chem. Phys.
  Lett.}\ }\textbf {\bibinfo {volume} {409}},\ \bibinfo {pages} {295} (\bibinfo
  {year} {2005})}\BibitemShut {NoStop}%
\bibitem [{\citenamefont {Wolf}, \citenamefont {Reiher},\ and\ \citenamefont
  {Hess}(2002)}]{wolf02}%
  \BibitemOpen
  \bibfield  {author} {\bibinfo {author} {\bibfnamefont {A.}~\bibnamefont
  {Wolf}}, \bibinfo {author} {\bibfnamefont {M.}~\bibnamefont {Reiher}}, \ and\
  \bibinfo {author} {\bibfnamefont {B.~A.}\ \bibnamefont {Hess}},\ }\href@noop
  {} {\bibfield  {journal} {\bibinfo  {journal} {J. Chem. Phys.}\ }\textbf
  {\bibinfo {volume} {117}},\ \bibinfo {pages} {9215} (\bibinfo {year}
  {2002})}\BibitemShut {NoStop}%
\bibitem [{\citenamefont {Huber}\ and\ \citenamefont
  {Herzberg}(1979)}]{hube79}%
  \BibitemOpen
  \bibfield  {author} {\bibinfo {author} {\bibfnamefont {K.~P.}\ \bibnamefont
  {Huber}}\ and\ \bibinfo {author} {\bibfnamefont {G.}~\bibnamefont
  {Herzberg}},\ }\href@noop {} {\emph {\bibinfo {title} {Molecular Spectra and
  Molecular Structure: Constants of Diatomic Molecules}}}\ (\bibinfo
  {publisher} {Van Nostrand Reinhold, New York},\ \bibinfo {year}
  {1979})\BibitemShut {NoStop}%
\bibitem [{\citenamefont {Gendron}\ \emph {et~al.}(2014)\citenamefont
  {Gendron}, \citenamefont {Pritchard}, \citenamefont {Bolvin},\ and\
  \citenamefont {Autschbach}}]{gend14}%
  \BibitemOpen
  \bibfield  {author} {\bibinfo {author} {\bibfnamefont {F.}~\bibnamefont
  {Gendron}}, \bibinfo {author} {\bibfnamefont {B.}~\bibnamefont {Pritchard}},
  \bibinfo {author} {\bibfnamefont {H.}~\bibnamefont {Bolvin}}, \ and\ \bibinfo
  {author} {\bibfnamefont {J.}~\bibnamefont {Autschbach}},\ }\href@noop {}
  {\bibfield  {journal} {\bibinfo  {journal} {Inorg. Chem.}\ }\textbf {\bibinfo
  {volume} {53}},\ \bibinfo {pages} {8577} (\bibinfo {year}
  {2014})}\BibitemShut {NoStop}%
\bibitem [{\citenamefont {{Dunning Jr.}}(1989)}]{dunn89}%
  \BibitemOpen
  \bibfield  {author} {\bibinfo {author} {\bibfnamefont {T.~H.}\ \bibnamefont
  {{Dunning Jr.}}},\ }\href@noop {} {\bibfield  {journal} {\bibinfo  {journal}
  {J. Chem. Phys.}\ }\textbf {\bibinfo {volume} {90}},\ \bibinfo {pages} {1007}
  (\bibinfo {year} {1989})}\BibitemShut {NoStop}%
\bibitem [{\citenamefont {Dyall}(2007)}]{dyal07b}%
  \BibitemOpen
  \bibfield  {author} {\bibinfo {author} {\bibfnamefont {K.~G.}\ \bibnamefont
  {Dyall}},\ }\href@noop {} {\bibfield  {journal} {\bibinfo  {journal} {Theor.
  Chem. Acc.}\ }\textbf {\bibinfo {volume} {117}},\ \bibinfo {pages} {491}
  (\bibinfo {year} {2007})}\BibitemShut {NoStop}%
\end{thebibliography}

\newcommand{\Aa}[0]{Aa}

\end{document}